\documentclass[sigplan,screen]{acmart}

\copyrightyear{2025} 
\acmYear{2025} 
\setcopyright{cc}
\setcctype[4.0]{by}
\acmConference[PPoPP '25]{The 30th ACM SIGPLAN Annual Symposium on Principles and Practice of Parallel Programming}{March 1--5, 2025}{Las Vegas, NV, USA}
\acmBooktitle{The 30th ACM SIGPLAN Annual Symposium on Principles and Practice of Parallel Programming (PPoPP '25), March 1--5, 2025, Las Vegas, NV, USA}
\acmDOI{10.1145/3710848.3710870}
\acmISBN{979-8-4007-1443-6/25/03}

\usepackage{subcaption}
\def\BibTeX{{\rm B\kern-.05em{\sc i\kern-.025em b}\kern-.08em
    T\kern-.1667em\lower.7ex\hbox{E}\kern-.125emX}}

\usepackage{amsmath,amsfonts}

\usepackage{graphicx}
\usepackage{textcomp}

\usepackage{hyperref}
\usepackage[normalem]{ulem}
\usepackage{tikz}
\usepackage{enumitem}
\usepackage{xspace}
\usepackage{multirow}
\usepackage{comment}
\usepackage{tcolorbox}
\usepackage{etoolbox}
\usepackage{multicol}
\usepackage{threeparttable}
\usepackage{xstring}
\usepackage{booktabs} 
\usepackage{slashbox}
\usepackage{wrapfig}
\usepackage{algcompatible}
\usepackage[ruled,linesnumbered]{algorithm2e}

\usepackage{bashful}
\usepackage{listings}

\newcommand{\ours}{ATTNChecker\xspace}
\newcommand{\att}{attention mechanism\xspace}
\newcommand{\fff}{INF, NaN, and near-INF\xspace}
\newcommand{\fffor}{INF, NaN, or near-INF\xspace}
\newcommand{\eec}{EEC-ABFT\xspace}

\newcommand{\mm}{GEMM\xspace}
\newcommand{\mms}{GEMMs\xspace}

\begin{document}

\title{\ours: Highly-Optimized Fault Tolerant Attention for Large Language Model Training}


\author{Yuhang Liang}
\affiliation{
    \institution{University of Oregon}
    \state{Oregon}
    \country{USA}
}
\email{yuhangl@uoregon.edu}

\author{Xinyi Li}
\affiliation{
    \institution{Pacific Northwest National Laboratory}
    \state{Washington}
    \country{USA}
}
\email{xinyi.li@pnnl.gov}

\author{Jie Ren}
\affiliation{
    \institution{College of William \& Mary}
    \state{Virginia}
    \country{USA}
}
\email{jren03@wm.edu}

\author{Ang Li}
\affiliation{
    \institution{Pacific Northwest National Laboratory}
    \state{Washington}
    \country{USA}
}
\email{ang.li@pnnl.gov}

\author{Bo Fang*}
\affiliation{
    \institution{Pacific Northwest National Laboratory}
    \state{Washington}
    \country{USA}
}
\email{bo.fang@pnnl.gov}

\author{Jieyang Chen*}
\affiliation{
    \institution{University of Oregon}
    \state{Oregon}
    \country{USA}
}
\email{jieyang@uoregon.edu}

\thanks{*Co-corresponding authors.}

\begin{abstract}

Large Language Models (LLMs) have demonstrated remarkable performance in various natural language processing tasks. However, the training of these models is computationally intensive and susceptible to faults, particularly in the attention mechanism, which is a critical component of transformer-based LLMs. 
In this paper, we investigate the impact of faults on LLM training, focusing on INF, NaN, and near-INF values in the computation results with systematic fault injection experiments. We observe the propagation patterns of these errors, which can trigger non-trainable states in the model and disrupt training, forcing the procedure to load from checkpoints.
To mitigate the impact of these faults, we propose \ours, the first Algorithm-Based Fault Tolerance (ABFT) technique tailored for the attention mechanism in LLMs. 
\ours is designed based on fault propagation patterns of LLM and incorporates performance optimization to adapt to both system reliability and model vulnerability while providing lightweight protection for fast LLM training. 
Evaluations on four LLMs show that \ours incurs on average 7\% overhead on training while detecting and correcting all extreme errors. Compared with the state-of-the-art checkpoint/restore approach, \ours reduces recovery overhead by up to 49$\times$.

\end{abstract}

\ccsdesc[500]{Computing methodologies~Machine learning}
\ccsdesc[500]{Computer systems organization~Parallel architectures}
\ccsdesc[500]{Computer systems organization~Dependable and fault-tolerant systems and networks}

\keywords{Algorithm-Based Fault Tolerance, Attention Mechanism, Large Language Models, Matrix Multiplication}

\maketitle

\section{Introduction} 
\label{sec:introduction}


Large Language Models (LLMs) commonly consume significant resources for training and fine-tuning. With the model consisting of more parameters and complexity, managing and updating this vast number of parameters during the training process demands substantial computational power. In addition, LLMs are trained on extensive collections of data, 
also consuming significant computing resources to finish the training steps. 
As a result, training an LLM is an extremely time and resource-intensive process, often taking weeks or even months to reach the desired accuracy. Most LLMs, therefore, are accelerated by high-performance systems featuring a massive number of GPUs, dedicated interconnects, and so on. For example, training GPT-3~\cite{brown2020language}, 
reportedly took thousands of GPUs. The MegaScale project~\cite{jiang2024megascale} indicates using 10,000 GPUs to meet the demand of such large-scale training. Smaller models like BERT~\cite{devlin2018bert}, which has around 110 million parameters for its base version, might require fewer resources but are still GPU-intensive tasks. 

As systems scale up and GPUs are extensively utilized, it is risky to experience system-wise transient hardware faults, a phenomenon well-documented in the field report of large systems~\cite{largescale_gpu, raditation-induced, extremescaleresilience, dixit2022detectingsilentdatacorruptions}. Training LLMs with HPC resources needs to overcome the soft errors. 
The growing size and complexity of LLMs make it imperative to mitigate the impact of soft errors during training. Previous studies~\cite{zhang2019quantifying} have explored the resilience of machine learning model training against minor data deviations caused by soft errors, concluding that such deviations often result in benign outcomes, with models typically converging successfully despite these errors. A more recent investigation~\cite{he2023understanding} provides a comprehensive analysis through systematic fault injection experiments conducted during model training. This study is particularly interesting as it offers insights into the implications of specific faults, notably Infinity (INF) and Not-a-Number (NaN) errors, in the context of LLM training. Unlike normal errors, LLM training typically cannot recover from INF or NaN errors via additional training cycles. 

The best practice to handle the occurrence of INF and NaN during the training is for the model to roll back to a checkpoint where no INF and NaN are present and retrain the model from the stored parameters~\cite{wang2023reliable}. While effective, retraining is considered a costly recovery method for LLM training in terms of time overhead, resource occupation, and expense in dollars~\cite{wu2023transom, wan2024bytecheckpoint,rajbhandari2020zero}. 
Moreover, since INF and NaN can propagate during the computation, it might be necessary for the model to roll back to an earlier checkpoint that is steps away, which further exacerbates the repair cost.  

To overcome the outcome of INF and NAN on the model training and resolve the inefficiency introduced by the current solution, we introduce an innovative algorithm-based fault tolerance technique (a.k.a ABFT) optimized for LLM training. The proposed technique, denoted as \ours, is designed to efficiently and effectively identify and rectify abnormal values in real-time during training. The primary objective of \ours is to develop a mechanism that is both lightweight and specifically tailored to the architectural needs of LLMs. \ours offers two advantages: First, the ability to detect anomalies where they initially occur prevents the further spread of errors through the system, thereby maintaining the integrity of the training process. Second, ABFT's capacity for immediate correction enables the system to recalibrate any infinite (INF), undefined (NaN), or other types of errors that can eventually lead to INF or NAN (i.e., near-infinite (near-INF) values) through the fast correction process. This not only relaxes the need for extensive rollback procedures but also significantly reduces the dependency on checkpoint-based recovery methods, thereby streamlining the training process and enhancing the overall training efficiency.

Several challenges need to be addressed for \ours to apply to LLMs, including: (a). Unlike normal numeric errors, \fff would corrupt the entire checksums, which is the key technique employed in most ABFT techniques for error detection and correction. The corrupted checksums are no longer capable of locating errors and recovering the correct values; (b). It is unclear to identify the particular operations for applying ABFT, as the LLMs generally consist of distinct network architectures. A universal application of ABFT techniques might cause unnecessary computation cycles and inefficient protection; (c). Different mathematical operations usually require dedicated ABFT techniques. Given such complex operational space in LLMs, it is unclear which particular ABFT techniques are more beneficial, while a trade-off between the scope of ABFT and optimization effort should be explored.

\ours aims to address the above challenges with the following assumptions. The attention mechanisms, particularly self-attention in Transformer-based LLMs, are among the most computationally demanding components in these models. This high demand stems from the quadratic increase in computational complexity with the length of input sequences, intensive matrix multiplication operations, and the large number of parameters within the attention layers. Furthermore, the attention mechanism's frequent and intensive memory access can heavily occupy the system's memory bandwidth. These factors combined make the attention mechanism a central focus not only for the purpose of optimizing the efficiency of LLMs but also for improving the overall fault tolerance of LLM training. 

Our paper makes the following contributions.
\begin{itemize}[leftmargin=*]
\renewcommand{\labelitemi}{\scriptsize$\blacksquare$}
    \item We conduct the first comprehensive fault injection and error propagation study on the \att against \fff errors. 
    In addition, we conduct the first vulnerability study for the key operations in the \att given \fff errors.
    \item We design the first ABFT that can effectively handle \fff errors - Extreme Error Correcting ABFT (\eec). \eec is also the first ABFT that is highly optimized to handle errors in various circumstances, including propagated errors, unpredictable patterns, and mixed error types.
    \item Based on \eec, we build the first comprehensive soft error protection approach for \att - \ours. \ours is specially tailored for protecting all major operations in attention and optimized for adaptive system reliability. 
    As \ours is completely transparent to models, any existing LLMs can improve reliability by integrating our \ours with minimum modification efforts and complementary with existing fault tolerance mechanisms. 
    \item We integrate \ours into the PyTorch framework. The evaluation shows that \ours incurs on average 7\% overhead on training while reaching 100\% detection and correction rate across all extreme errors. Our performance estimation shows that 
    \ours reduces recovery overhead by up to $49\times$ compared with the state-of-the-art recovery technique.  
\end{itemize}

\section{Background} \label{sec:background}

\subsection{Attention in LLMs}
The attention mechanism is the core building block in Large Language Models (LLMs).  It maps a query and a set of key-value pairs to the result of a weighted sum, effectively providing a contextually relevant representation based on the input query and the key-value pairs~\cite{vaswani2017attention}. Query, Key, and Value are matrices, each representing instance vectors. The Self-attention could be described as:
\begin{equation} \label{attntion}
    Attention(Q,K,V) = softmax(\frac{QK^T}{\sqrt{d_{k}}})V 
\end{equation} where $d_{k}$ is the dimension of Q and K, and $\sqrt{d_{k}}$ used to scale the result of $AS=QK^T$ (called Attention Scores).

\begin{figure}[h!]
    \vspace{-1em}
    \centering
    \includegraphics[width=0.4\textwidth]{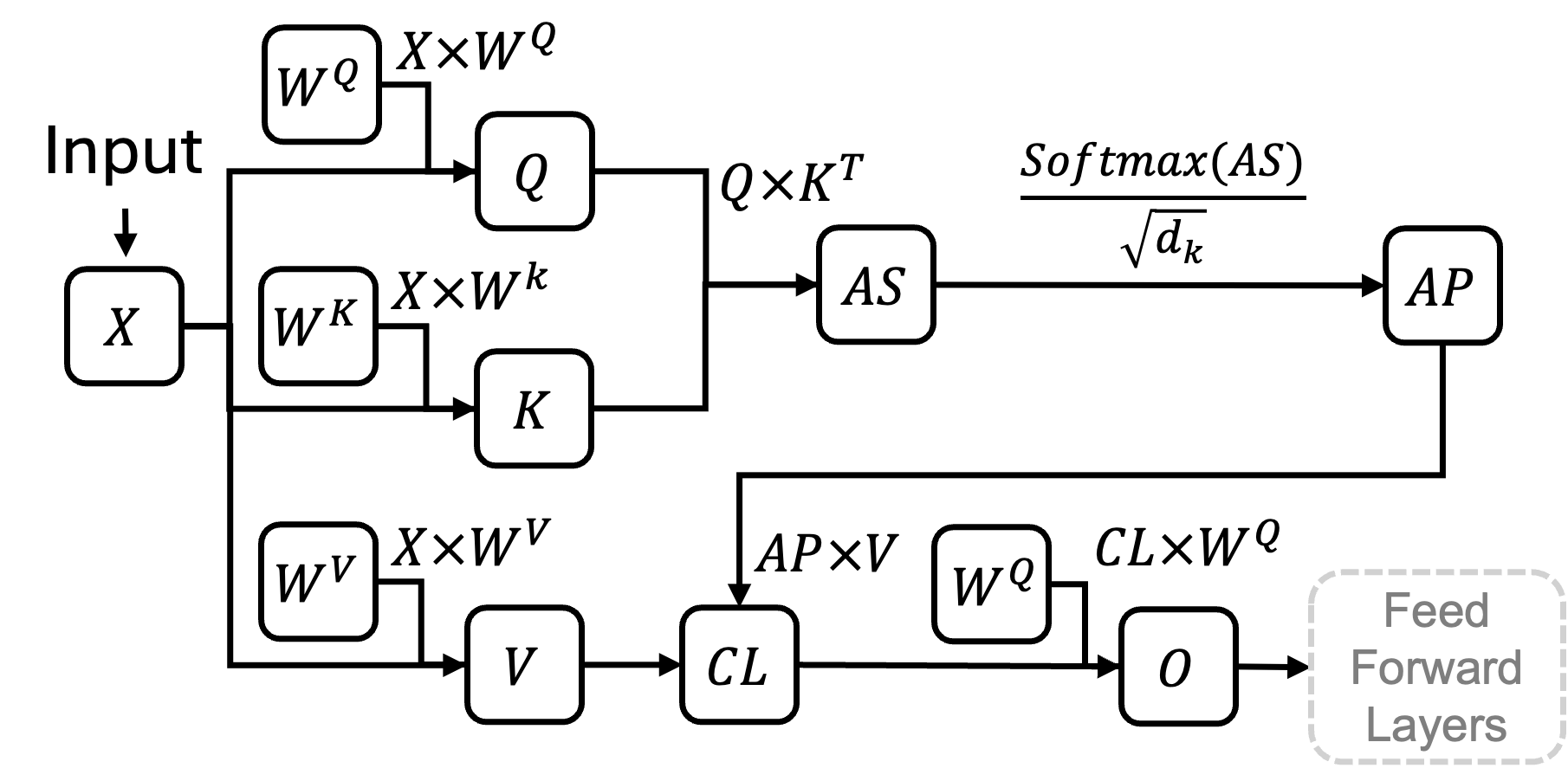}
    \vspace{-1em}
    \caption{Execution Flow in Attention}
    \label{fig:AttnFlow}
    \vspace{-1.5em}
\end{figure}

Based on self-attention, the multi-head attention\cite{vaswani2017attention} is implemented by linearly projecting Q, K, and V multiple times, then each head performs self-attention based on its own Q, K, and V. The results of all heads will be combined. Multi-Head Attention is described as:
\begin{multline} \label{multihead}
    MultiHead(Q,K,V) = Concat(head_{1}, head_{2}, ... , head_{n})W^{o}\\
    each\, head_{i} = Attention(QW_i^q, KW_i^k, VW_i^v)
\end{multline} 
All matrix operations in the multi-head attention can be described using the execution flow shown in Figure~\ref{fig:AttnFlow} including six matrix multiplications and one softmax operation.

\subsection{Soft Errors causing \fffor Values}
According to IEEE754, the binary representation of floating point numbers includes 3 basic parts: sign, exponent, and mantissa. 
Depending on where bit-flips occur they can affect the original value differently.  
Two specific types of floating-point errors can be caused by bit-flips: INF (i.e., infinity) and NaN (i.e., Not-a-Number).
Although not classified as an exception by convention, extremely large values (near-INF) resulting from bit-flips in the exponent can also negatively impact computation.
There are no specific rules for identifying near-INF, we generally consider values that are larger than a threshold as near-INF, which can vary for different computations.
\fff could cause several exceptions during the computation, as they can legally propagate through calculations e.g., INF times any number equals INF.
In addition, one type of exception can transit to another type e.g., A near-INF number operating on another near-INF may lead to INF.

\begin{figure}[t]
    \vspace{-1em}
    \includegraphics[width=1\columnwidth]{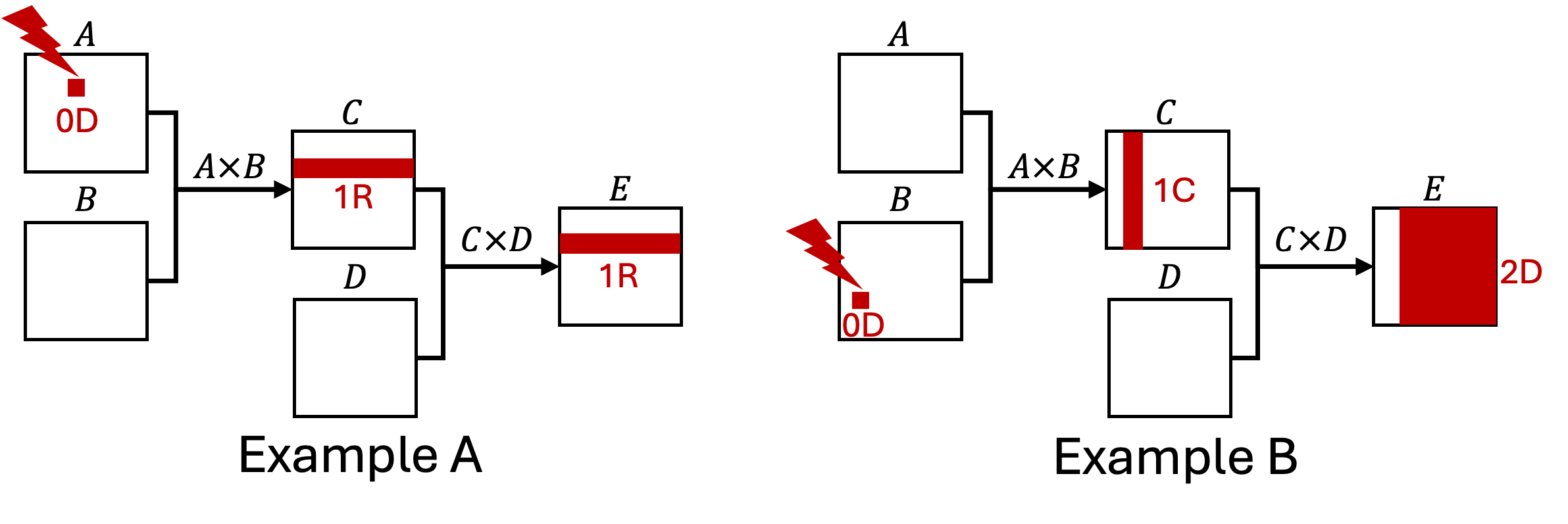}
    \vspace{-2.5em}
    \caption{Two Examples of Propagation in GEMM.  (A): Error Only Propagates to 1R. (B): Error Propagates to 1C then 2D.}
    \label{error-pattern}
    \vspace{-1em}
\end{figure}

In a matrix multiplication (GEMM) $C = A \times B$, where $A$ and $B$ are input and $C$ is the output, the presence of error elements can negatively impact the result through propagation.
Figure~\ref{error-pattern} shows two examples of error propagations in GEMM. The propagation patterns can be classified as\cite{chen2018fault}:
\begin{itemize}
    \item 0D: One standalone error without propagation. It typically appears in the origin matrix where fault strikes.
     \item 1D: Errors accumulate along one row or column (entire or partial). We distinguish them using 1R for one row and 1C for one column.
     \item 2D: Errors accumulate beyond 1D such as a sub-matrix.
\end{itemize}

\subsection{ABFT for Matrix Multiplication}



ABFT~\cite{huang1984algorithm} uses redundant information to verify the correctness of a particular algorithm under the presence of errors. For matrix operations, ABFT encodes redundant information using checksums for error protection~\cite{chen2008algorithm,chen2018fault}.
To apply ABFT on GEMM, input matrix $A$ and $B$ are first encoded with column checksum $A^c$ and row checksum $B^r$, where a checksum is a (weighted) sum of matrix elements along columns/rows.
To both detect and correct errors, two checksums are needed.
A typical choice of the two checksum weights are: $v_1  = {[1,1,1...1]}^T$ (unweighted) and $v_2  = {[1,2,3...n]}^T$ (weighted).
So the column and row checksums can be encoded as $A^c= \begin{bmatrix}
{v_1}^T
\\
{v_2}^T
\end{bmatrix}\cdot A$
and 
$B^r=B\cdot \begin{bmatrix}
v_1
v_2
\end{bmatrix}$.
Next, we compute the output $C$ and its checksums:
$
C^c = A^c \cdot B
$ and
$
C^r = B \cdot A^r
$
To detect errors, one needs to check if the linear relationship still holds for matrix $C$ by recalculating the checksums ($C^{c'}$ and $C^{r'}$) and comparing with $C^{c}$ and $C^{r}$. 
Using the column checksum correction as an example (row checksum correction is similar): we compare $C^{c'}$ with $C^{c}$ to see whether they are close enough (within roundoff error $E$ ~\cite{chen2018fault}) by calculating: $\delta = C^{c} - C^{c'}$.
For instance, if $\left  |\delta_{0,j}\right |> E$, then an error is detected on the $j^{th}$ column of $C$. 
Then, $\delta_{1,j}/\delta_{0,j}=i$ (round to the nearest integer) calculates the error's row index $i$ and $\delta_{0,j}$ is the difference between the true and corrupted value, which can be used for correction.




Existing ABFTs\cite{chen2016online, chen2018fault, wu2016towards, chen2016gpu, wu2017silent, liang2017correcting} are designed to handle moderate value changes due to errors. 
However, they cannot efficiently handle extreme data corruptions that lead to INF, NaN, or near-INF values.
This is because when locating the error, $\delta_{1,*}/\delta_{0,*}$ would be INF or NaN if the error is INF or NaN, causing incorrect error index calculation.
Although not guaranteed, an error with a value close to INF (i.e., near-INF) might also lead to INF in $\delta$ due to the overflow.
Given such limitations, existing ABFTs are not suitable to detect and correct errors that lead to INF, NaN, or near-INF values. 

\begin{table}[t]
    \vspace{-1em}
    \centering
    \caption{Table of Notation}
    \vspace{-1em}
    \begin{tabular}{|c|c|}
        \hline 
        \textbf{Notation} & \textbf{Description}\\
        \hline 
        $Q$ & Query \\
        \hline 
        $K$ & Key\\
        \hline
        $V$ & Value\\
        \hline
        $AS$ & Attention Score\\
        \hline
        $AP$ & Attention Probability\\
        \hline
        $CL$ & Context Layer\\
        \hline
        $O$ & Output \\
        \hline
    \end{tabular}
    \label{tab:notations}
    \vspace{-1em}
\end{table}

\section{Fault Injection and Error Propagation Study for Attention} 
\label{sec:fault-injection}
\begin{table}[t]
\caption{Error Propagation Patterns in Attention Mechanism. FI: fault injecting matrix. $\infty$/$\Theta$/N: INF, NaN, and near-INF. $\infty^*$: mixture of $+/-\infty$. $M$: mixture of all fault types.}
\vspace{-0.5em}
\begin{tabular}{cc|ccccccc|}
\cline{3-9}
\multicolumn{1}{l}{}                             & \multicolumn{1}{l|}{} & \multicolumn{7}{c|}{$\underrightarrow{\text{Propagating\ Matrix}}$}                                                                                                                                                                    \\ \cline{3-9} 
\multicolumn{1}{l}{}                             &                       & \multicolumn{1}{c|}{Q}      & \multicolumn{1}{c|}{K}      & \multicolumn{1}{c|}{V}      & \multicolumn{1}{c|}{AS}         & \multicolumn{1}{c|}{AP}         & \multicolumn{1}{c|}{CL}         & O          \\ \hline
\multicolumn{1}{|c|}{\multirow{5}{*}{\rotatebox{90}{Inject INF($\infty$)}}} & Q                     & \multicolumn{1}{c|}{\textbf{FI}} & \multicolumn{1}{c|}{-}      & \multicolumn{1}{c|}{-}      & \multicolumn{1}{c|}{1R-$\infty$*} & \multicolumn{1}{c|}{1R-$\Theta$} & \multicolumn{1}{c|}{1R-$\Theta$} & 1R-$\Theta$ \\ \cline{2-9} 
\multicolumn{1}{|c|}{}                           & K                     & \multicolumn{1}{c|}{}       & \multicolumn{1}{c|}{\textbf{FI}} & \multicolumn{1}{c|}{-}      & \multicolumn{1}{c|}{1C-$\infty$*} & \multicolumn{1}{c|}{2D-$\Theta$}     & \multicolumn{1}{c|}{2D-$\Theta$}     & 2D-$\Theta$     \\ \cline{2-9} 
\multicolumn{1}{|c|}{}                           & V                     & \multicolumn{1}{c|}{}       & \multicolumn{1}{c|}{}       & \multicolumn{1}{c|}{\textbf{FI}} & \multicolumn{1}{c|}{-}          & \multicolumn{1}{c|}{-}          & \multicolumn{1}{c|}{1C-M}   & 2D-M   \\ \cline{2-9} 
\multicolumn{1}{|c|}{}                           & AS                    & \multicolumn{1}{c|}{}       & \multicolumn{1}{c|}{}       & \multicolumn{1}{c|}{}       & \multicolumn{1}{c|}{\textbf{FI}}     & \multicolumn{1}{c|}{1R-$\Theta$} & \multicolumn{1}{c|}{1R-$\Theta$} & 1R-$\Theta$ \\ \cline{2-9} 
\multicolumn{1}{|c|}{}                           & CL                    & \multicolumn{1}{c|}{}       & \multicolumn{1}{c|}{}       & \multicolumn{1}{c|}{}       & \multicolumn{1}{c|}{}           & \multicolumn{1}{c|}{}           & \multicolumn{1}{c|}{\textbf{FI}}     & 1R-$\infty$ \\ \hline

\multicolumn{1}{|c|}{\multirow{5}{*}{\rotatebox{90}{Inject NaN($\Theta$)}}} & Q                     & \multicolumn{1}{c|}{\textbf{FI}} & \multicolumn{1}{c|}{-}      & \multicolumn{1}{c|}{-}      & \multicolumn{1}{c|}{1R-$\Theta$} & \multicolumn{1}{c|}{1R-$\Theta$} & \multicolumn{1}{c|}{1R-$\Theta$} & 1R-$\Theta$ \\ \cline{2-9} 
\multicolumn{1}{|c|}{}                           & K                     & \multicolumn{1}{c|}{}       & \multicolumn{1}{c|}{\textbf{FI}} & \multicolumn{1}{c|}{-}      & \multicolumn{1}{c|}{1C-$\Theta$} & \multicolumn{1}{c|}{2D-$\Theta$}     & \multicolumn{1}{c|}{2D-$\Theta$}     & 2D-$\Theta$     \\ \cline{2-9} 
\multicolumn{1}{|c|}{}                           & V                     & \multicolumn{1}{c|}{}       & \multicolumn{1}{c|}{}       & \multicolumn{1}{c|}{\textbf{FI}} & \multicolumn{1}{c|}{-}          & \multicolumn{1}{c|}{-}          & \multicolumn{1}{c|}{1C-$\Theta$}   & 2D-$\Theta$   \\ \cline{2-9} 
\multicolumn{1}{|c|}{}                           & AS                    & \multicolumn{1}{c|}{}       & \multicolumn{1}{c|}{}       & \multicolumn{1}{c|}{}       & \multicolumn{1}{c|}{\textbf{FI}}     & \multicolumn{1}{c|}{1R-$\Theta$} & \multicolumn{1}{c|}{1R-$\Theta$} & 1R-$\Theta$ \\ \cline{2-9} 
\multicolumn{1}{|c|}{}                           & CL                    & \multicolumn{1}{c|}{}       & \multicolumn{1}{c|}{}       & \multicolumn{1}{c|}{}       & \multicolumn{1}{c|}{}           & \multicolumn{1}{c|}{}           & \multicolumn{1}{c|}{\textbf{FI}}     & 1R-$\Theta$ \\ \hline

\multicolumn{1}{|c|}{\multirow{5}{*}{\rotatebox{90}{Inject nINF(N)}}} & Q                     & \multicolumn{1}{c|}{\textbf{FI}} & \multicolumn{1}{c|}{-}      & \multicolumn{1}{c|}{-}      & \multicolumn{1}{c|}{1R-M} & \multicolumn{1}{c|}{1R-$\Theta$} & \multicolumn{1}{c|}{1R-$\Theta$} & 1R-$\Theta$ \\ \cline{2-9} 
\multicolumn{1}{|c|}{}                           & K                     & \multicolumn{1}{c|}{}       & \multicolumn{1}{c|}{\textbf{FI}} & \multicolumn{1}{c|}{-}      & \multicolumn{1}{c|}{1C-M} & \multicolumn{1}{c|}{2D-$\Theta$}     & \multicolumn{1}{c|}{2D-$\Theta$}     & 2D-$\Theta$     \\ \cline{2-9} 
\multicolumn{1}{|c|}{}                           & V                     & \multicolumn{1}{c|}{}       & \multicolumn{1}{c|}{}       & \multicolumn{1}{c|}{\textbf{FI}} & \multicolumn{1}{c|}{-}          & \multicolumn{1}{c|}{-}          & \multicolumn{1}{c|}{1C-M}   & 2D-M   \\ \cline{2-9} 
\multicolumn{1}{|c|}{}                           & AS                    & \multicolumn{1}{c|}{}       & \multicolumn{1}{c|}{}       & \multicolumn{1}{c|}{}       & \multicolumn{1}{c|}{\textbf{FI}}     & \multicolumn{1}{c|}{1R-M} & \multicolumn{1}{c|}{1R-M} & 1R-M \\ \cline{2-9} 
\multicolumn{1}{|c|}{}                           & CL                    & \multicolumn{1}{c|}{}       & \multicolumn{1}{c|}{}       & \multicolumn{1}{c|}{}       & \multicolumn{1}{c|}{}           & \multicolumn{1}{c|}{}           & \multicolumn{1}{c|}{\textbf{FI}}     & 1R-M \\ \hline
\end{tabular}
\label{PropagationTab}
\end{table}

Prior studies~\cite{he2023understanding} have shown that a small perturbation affecting the LLM training data is likely to be mitigated over the model convergence process, and the final model's accuracy would be pertained under such errors. However, in terms of floating point values that belong to special categories like \fff, prior studies have demonstrated that the LLM models would be seriously compromised under such errors, leading to unacceptable outcomes. Moreover, a near-INF number is also problematic as it can either accumulate and result in INF/NaN or disrupt the gradient distribution to reduce the convergence probability. 

To understand the different impacts of soft errors on the LLM model training process, we conduct a fault injection and error propagation study on four representative LLMs: Bert, GPT-2, GPT-Neo, and Roberta. We specifically focus on studying the impact of faults that lead to \fff since those lead to the most drastic value change that may greatly impact model training.
The goal of our study is to understand (1) how an error propagates across operations in attention modules; 
(2) How do such errors impact the training state? We hypothesize that the answers to these questions would guide the design of efficient and effective ABFT approaches for attention modules.

We select the compute-intensive \mms in the \att as the fault injection site since they consume most of the computation cycles for the entire training process. (Table~\ref{tab:gemm-in-attention} shows the workload ratios of \mm in \att in four LLMs~\cite{hoffmann2022training}.)
In particular, we inject one fault (i.e., 0D pattern) into the result matrix of each \mm to simulate a fault that occurred during the computation. Then, we carefully examine the errors in matrices involved in the following operations. The notations are presented in Table \ref{tab:notations}
\begin{table}[t]
    \vspace{-1em}
    \centering
    \caption{Ratios of Matrix Multiplications Workloads Relative to the Entire Attention Mechanism}
    \vspace{-1em}
    \begin{tabular}{|c|c|c|c|c|}
        \hline 
        \textbf{Model} & Bert & GPT-2 & GPT-Neo & Roberta \\
        \hline 
        \textbf{GEMM Ratio} & 99.7\% & 99.5\% & 99.3\% & 99.7\%\\
        \hline 
    \end{tabular}
    \label{tab:gemm-in-attention}
    \vspace{-1em}
\end{table}

\subsection{Observation summary}
Table \ref{PropagationTab} summarizes the error propagation patterns in the downstream matrices following each fault injecting matrices: $Q$, $K$, $V$, $AS$, and $CL$ inside the \att.
We combine the discovery from all four LLMs as the error patterns are similar. We observe that errors can quickly propagate with diverse types of chances throughout the execution attention.
In addition, we evaluate the error's impact on the training status (e.g., training loss) and compare them with fault-free executions. As the studied space is extensive, we ensure our findings are statistically significant by repeatedly injecting each type of fault. Specifically, for each \mm, we randomly select 10\% ($\sim$5,000) elements of each output matrix to increase the statistic confidence and present our findings collectively. 
We mainly focus on whether an error will cause a non-trainable state: a numerical data corruption that causes a loss being NaN (i.e. we only observe that the loss being NaN would lead to an untrainable state).
Table~\ref{StatTab} shows the probability of a non-trainable state caused by different error types. 
We observe very diverse fault sensitivity across different operations in attention.

\begin{table}[t]
    \captionsetup{justification=centering}
    \caption{Probability of Non-trainable States Caused by Error}
    \vspace{-0.5em}
    \begin{tabular}{|c|c|c|c|c|c|c|}
    \hline
    Error&\multirow{2}{*}{Model}&\multicolumn{5}{|c|}{Fault Injecting Matrix} \\
    \cline{3-7} 
    Type & \textbf{} & Q& K& V & AS  & CL\\
    \hline
    \multirow{4}{*}{INF}   
                &Bert&100\% &100\% &100\% &100\% &100\%\\
    \cline{2-7}
                &GPT-2&91.8\% &86.8\% &100\% &56.9\% &100\%\\
    \cline{2-7}
                &Neo&100\% &85.6\% &100.0\% &54.7\% &100\%\\
    \cline{2-7}
                &Roberta&100\% &99.9\% &100.0\% &100\% &100\%\\
    \hline

    \multirow{4}{*}{NaN}   
                &Bert &100\% &100\% &100\% &100\% &100\%\\
    \cline{2-7}
                &GPT-2 &100\% &100\% &100\% &54.7\% &100\%\\
    \cline{2-7}
                &Neo &100\% &100\% &100.0\% &54.7\% &100\%\\
    \cline{2-7}
                &Roberta &100\% &100\% &100.0\% &100\% &100\%\\
    \hline

    \multirow{4}{*}{nINF}   
                &Bert&45.9\% &43.4\% &6.3\% &0.2\% &0.6\%\\
    \cline{2-7}
                &GPT-2&38.4\% &37.2\% &1.0\% &0.5\% &0.7\%\\
    \cline{2-7}
                &Neo &10.3\% &14.4\% &5.8\% &11.2\% &9.6\%\\
    \cline{2-7}
                &Roberta&54.0\% &49.9\% &3.6\% &5.5\%  &0.4\%\\
    \hline
    
    \end{tabular}
    \label{StatTab}
    \vspace{-1em}
\end{table}

\subsection{Insight for Designing Fault Tolerance}
Next, based on our error propagation study, we summarize our insights to help us design efficient and effective fault tolerance for \att in Section~\ref{sec:method}.

\underline{Segmented Protection}: 
As error propagations could easily lead to a catastrophic result, operations in \att need to be segmented with fault detection at the operation boundary to confine the error.
For example, if a 2D pattern is beyond recoverable, then error detection/correction needs to be done for $AS$, $CL$, and $O$ to minimize the probability of the occurrence of a 2D pattern.

\underline{Handling (Non) deterministic Patterns}: For operations with deterministic error patterns, efficient fault tolerance can be implemented as the deterministic pattern is bounded with no additional pattern detection required. For instance, if we use segmented fault tolerance, then $O$ will have only deterministic 1R patterns (0D pattern can be handled the same way as 1R). However, for operations with nondeterministic error patterns, fault tolerance needs to be adaptive to different potential patterns. For example, the error pattern of $AS$ can be either 1R or 1C, depending on the origin of the error.

\underline{Handling Mixed-type Patterns}: In many cases, the propagated pattern contains a mixture of error types (e.g., +/-INF or a mixture of three types). In those cases, error detection and correcting can be challenging because different error types need to be handled differently. In addition, the detection of one error type may cause the error to transit to another type. For example, computing for detecting near-INF errors may result in INF errors. So, fault tolerance needs to be specially designed to handle mixed-type patterns and type transitions.

\underline{Selective Error Handling}: The vulnerability of each operation varies across different error types and models. Higher vulnerable operation requires close attention (e.g., $Q$), while less vulnerable ones (e.g., $CL$) would likely benefit from less intensive fault tolerance methods. To balance both fault tolerance coverage and overhead, such methods must be selective against different types of errors and different models.

\section{ABFT for Attention Mechanism} 
\label{sec:method}

In this section, we propose to build the first ABFT for the \att, a key building block of many LLMs.
This section introduces our fault model, which defines the types of faults \ours is interested in.
Then, we introduce ABFT with an extreme error correcting algorithm (\eec) against \fff errors and various types of propagated errors in attention.
We then use \eec to build a systematic protection for the \att denoted as \ours.
In addition to providing comprehensive protection, \ours also makes ABFT detection adaptable to system reliability and model vulnerability.
Finally, we build highly-optimized GPU kernels for achieving lightweight ABFT protection for the \att.
Our goal is to maximize the coverage of \att against soft errors such that the reliability of LLM training can be greatly improved while the training performance is well-maintained. 






\subsection{Fault Model}

The occurrence of INF, NaN, and near-INF can originate from diverse sources: (1) Input-Related Faults: Certain inputs to the model can trigger floating-point exceptions during numerical operations, leading to INF/NaN values. (2) Hardware Faults: Bit-flips induced by radiation or manufacturing defects can affect the correctness of computations, generating extreme values. Note that memory is usually ECC-protected in HPC systems; hence, a memory error is beyond the scope of our paper. (3) Inappropriate Hyperparameters: The use of unsuitable hyperparameters, like an excessively large learning rate, as indicated in prior research~\cite{xu2021learning}, can cause a rapid increase in the loss function (known as loss explosion), resulting in INF/NaN values.

Input-related faults, likely caused by certain input data, cannot be effectively addressed using ABFT. This is because repeating the same computations during training would reproduce these exceptions. Conversely, issues related to loss explosion, often a consequence of inappropriate hyperparameters, can be dynamically mitigated through parameter adjustments. Therefore, the focus of the paper is to address the impact of transient hardware faults that lead to soft errors caused \fffor in the \att. These hardware-related errors are particularly problematic as they not only cause severe disruptions in the training process but also entail substantial overhead for correction in existing training frameworks.
As \att is one the most important building blocks of many LLMs, this work targets soft errors that occur during the execution of \att.

\begin{figure}[t]
    \vspace{-1em}
    \centering
    \includegraphics[width=0.9\columnwidth]{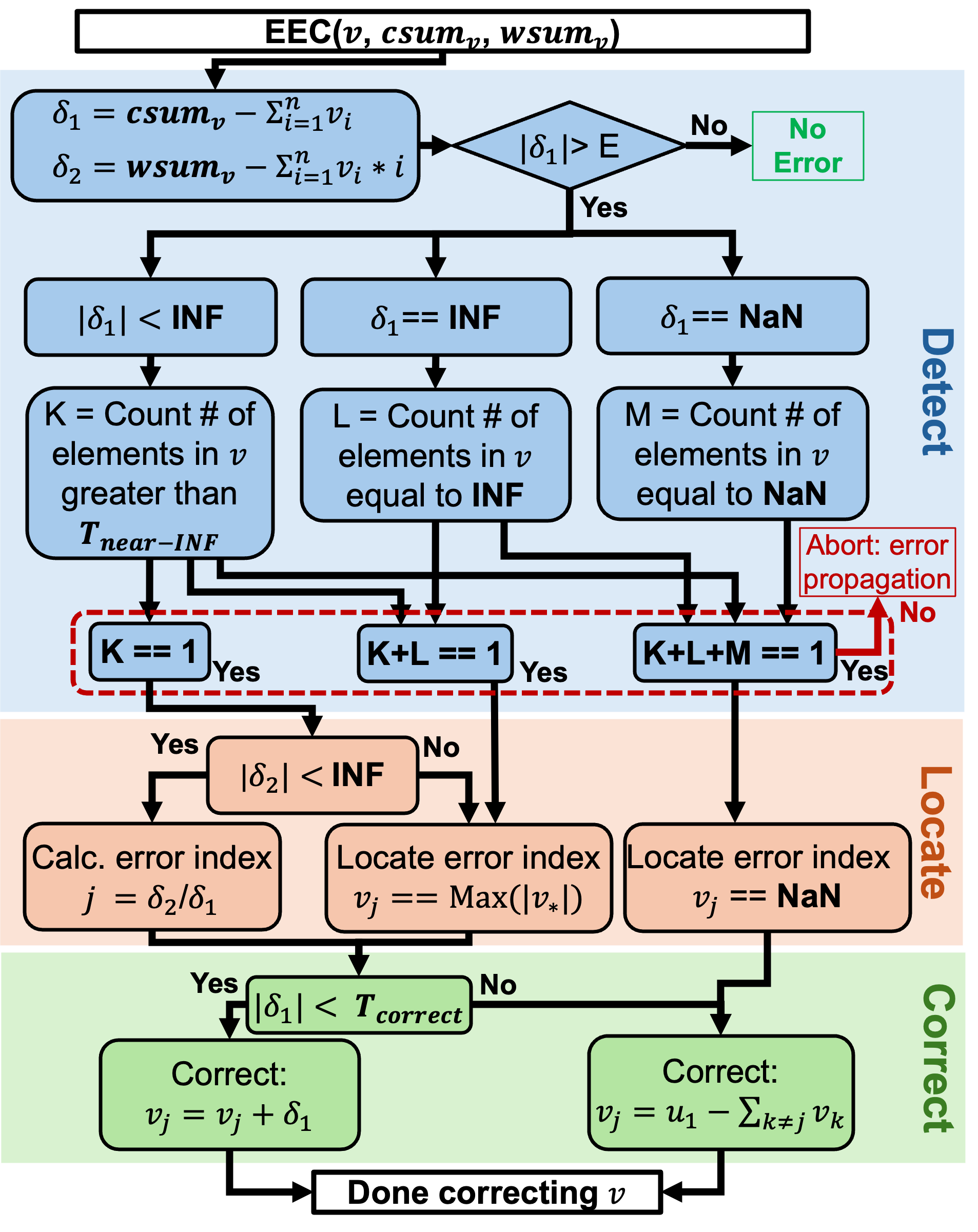}
    \vspace{-1em}
    \caption{Extreme Error Correcting ABFT Enabling Differentiated Error Handling for \fff Errors.}
    \label{Enhaced-ABFT}
    \vspace{-1em}
\end{figure}

\subsection{Extreme Errors Correcting ABFT}
\label{sec:eec}
Figure~\ref{Enhaced-ABFT} shows our extreme error correcting ABFT (\eec)  handling errors in one column or row vector $v$ in a matrix given the non-weighted checksum $csum_v$ and weighted checksum $wsum_v$.
Similar to existing ABFT,  \eec first compares the updated checksums $csum_v$ and $wsum_v$ with recalculated checksums to get the difference $\delta_1$ and $\delta_2$, then use $\delta_1$ to detect the error.
Upon detecting an error, instead of relying on a unified error handling strategy, \eec applies corresponding error locating and correction procedures for different cases:

\underline{Case 1}: $\delta_1 < INF$ indicates all error(s) are less than INF.
Then we count the number of elements that are greater than a threshold $T_{near-INF}$ for identifying near-INF.
If such a number $K$ is one, then it indicates that the error is not propagated.
Next, we precisely locate the error.
Normally, one can calculate the error index using $\delta_2/\delta_1$, but we must be careful since the weighted checksum of a vector containing near-INF elements can cause overflow in $\delta_2$.
So, \eec checks the value of $\delta_2$ before locating the error.
If $\delta_2 < INF$, we proceed with the regular locating logic.
Otherwise, we search for the largest value in $v$ to find the error index.
Finally, \eec corrects the error given the location.
Special attention is worth paying to error correction for near-INF cases since a round-off error may cause a corrupted value to be wrongly corrected. 
In this case, $\delta_1$ will absorb smaller $v_j$ in the correction process leading to incorrect recovery.
To avoid such precision-related errors, we define a threshold $T_{correct}$ to determine which error-correcting strategy to use. Empirically, we use $T_{near-INF} = 10^{10}$ and $T_{correct} = 10^5$.

\underline{Case 2}: If $\delta_{0,i} == INF$, then it is either an INF error in the output or a near-INF error that causes $\delta_1$ to overflow. 
As $\delta_2$ may overflow, we search the max value for locating the error. 
To correct the error, since overflow occurs to both $\delta_1$ and $\delta_2$, one cannot use them to directly perform the correction. 
In this case, \eec needs to reconstruct the corrupted value using existing values.

\underline{Case 3}: If $\delta_i == NaN$, then all three types of errors may occur since the NAN can be caused by INF or near-INF involved computation. 
To locate the error, \eec needs to search the corresponding index for NaN and use the same way as case 2 to reconstruct the corrupted value. 

\underline{Case 4}: 
For all the previous three cases, we check if there is an error propagation in $v$ by counting the number of errors. If it is greater than one, a 1D error propagation is identified, then \eec will abort the current correction process as we need to initiate a special procedure for error propagation. (will be discussed in Section~\ref{nonder}).

\subsection{Correcting Extreme Error Propagations}
Next, we describe how we adaptively handle three common error patterns in the \att and discuss how we efficiently parallelize the correction procedures on GPUs. 
Note that we limit \eec to handle up to 1D patterns for all cases since 2D patterns cannot be efficiently corrected.
We defer our discussion on propagation mitigation to Section~\ref{abft-scheme}, where we design systematic ABFT schemes for attention.

\begin{figure}[t]
    \vspace{-1em}
    \centering
    \includegraphics[width=0.9\columnwidth]{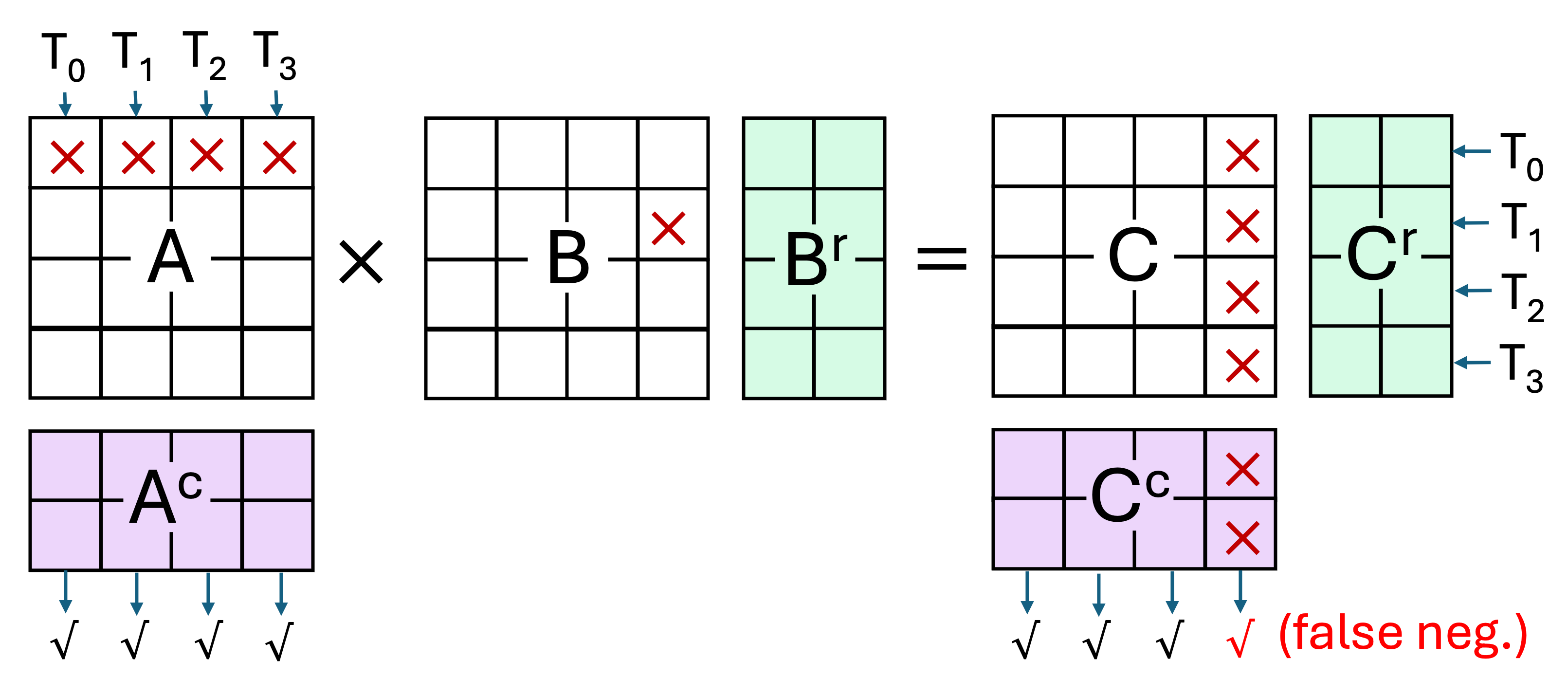}
    \vspace{-1em}
    \caption{Deterministic (1R in $A$) and Nondeterministic (1C in $C$) Error Pattern Handling Using \eec ($T_0$-$T_3$ represents four threads)}
    \label{PropagationHandling}
    \vspace{-1.5em}
\end{figure}

\textbf{\textit{Deterministic Patterns:}}
An error pattern is deterministic for an operation when only one type of propagation pattern can occur in its output.
In this case, we only need to maintain the corresponding checksums for error correcting.
For example, the 1C pattern only needs row checksums, and the 1R pattern only needs column checksums.
The 0D pattern can be corrected using either checksum.
Figure~\ref{PropagationHandling} shows error-correcting on two matrices: $A$ and $C$. Assuming $A$ has a deterministic 1R error, we illustrate how to detect such patterns efficiently in parallel using \eec.
Specifically, we let each thread execute \eec on column vector in parallel on GPU.
We use $T_0$ - $T_3$ to represent four threads for correcting $A$ as an example.
For fault-free execution, all executions are divergence-free on GPUs.
For the 0D pattern, only the error-handling thread will execute the correction procedure, and all other threads will exit early to release the GPU resources.
For 1D patterns, all threads will execute the correction procedure divergent-free when all error types are the same. Slight divergence may occur when a mixed-type pattern occurs.




\label{nonder}

\textbf{\textit{Nondeterministic Patterns:}} 
Nondeterministic patterns raise two challenges: (1) the pattern is unknown in advance; (2) when corrupted elements are used for the following operation, they may not only propagate the error to the output but also corrupt the checksums of the output. 
For the 1C pattern, column checksums will be corrupted. For the 1R pattern, on the other hand, the row checksums will be corrupted.
Figure~\ref{PropagationHandling} illustrates the process of recovering a nondeterministic 1D error with 1C being the true pattern and column checksum being corrupted (due to $C^c = A^c \times B$, assuming $A$ is error-free after correction).
We need to maintain both column and row checksums to handle such a pattern.
To recover, we first try using column checksums to fix potential errors.
We will encounter one of two cases: (1) If non-extreme errors occur: Since the column checksums are calculated using the corrupted value, they will wrongly hold a relationship with the corrupted $C$, so no error is detected (false negative). (2) if extreme errors occur: \eec will recognize and report the propagation (case 4 of Section~\ref{sec:eec}).
In either case, no error correction will be done and we will then proceed to use uncorrupted row checksums to fix the errors.
Finally, we will use the corrected columns to recover the corrupted column checksums using re-computation.
If a 1R pattern occurs, on the other hand, column checksums would be able to correct all errors in the first step.

\textbf{\textit{Mixed-type Patterns:}} 
When a mixed-type pattern occurs, we cannot directly conclude the error type based on the value $\delta_1$, because of error type transition. 
For example, 
near-INF in the 1D pattern can lead to $\delta_1$ being INF, which may cause confusion for error detection. 
\eec overcomes such a challenge by counting all possible error types before making conclusions about error patterns. For example, 
if $\delta_1 = INF$, we count both INF and near-INF. 
If $\delta_1 = NAN$, all three kinds of error need to be considered. 

\subsection{Systematic ABFT Scheme for Attention}
\label{abft-scheme}
Next, we describe how to use \eec to build systematic ABFT protection for attention.
As our \eec can correct errors up to 1D, we divide the whole execution flow of \att into three \textit{protection sections} to minimize the probability of more serious error propagations (e.g., 2D). Specifically, the three sections are: $S_{AS} = \{X \times W^Q, X \times W^K, Q\times K^T\}$, $S_{CL} = \{X \times W^V, AP \times V \}$, $S_{O} = \{CL \times W^O\}$.
In addition, to ensure propagated errors can be effectively corrected and reduce error detection overhead, we carefully design a checksum passing mechanism for each section.
Specifically, we only need to compute the initial checksums for the inputs of the first operation and pass the updated checksums between operations to form protection for a whole section.
This enables us to correct up to 1 error at a time per section.

\begin{figure}[t]
    \vspace{-1em}
    \centering
    \includegraphics[width=1\columnwidth]{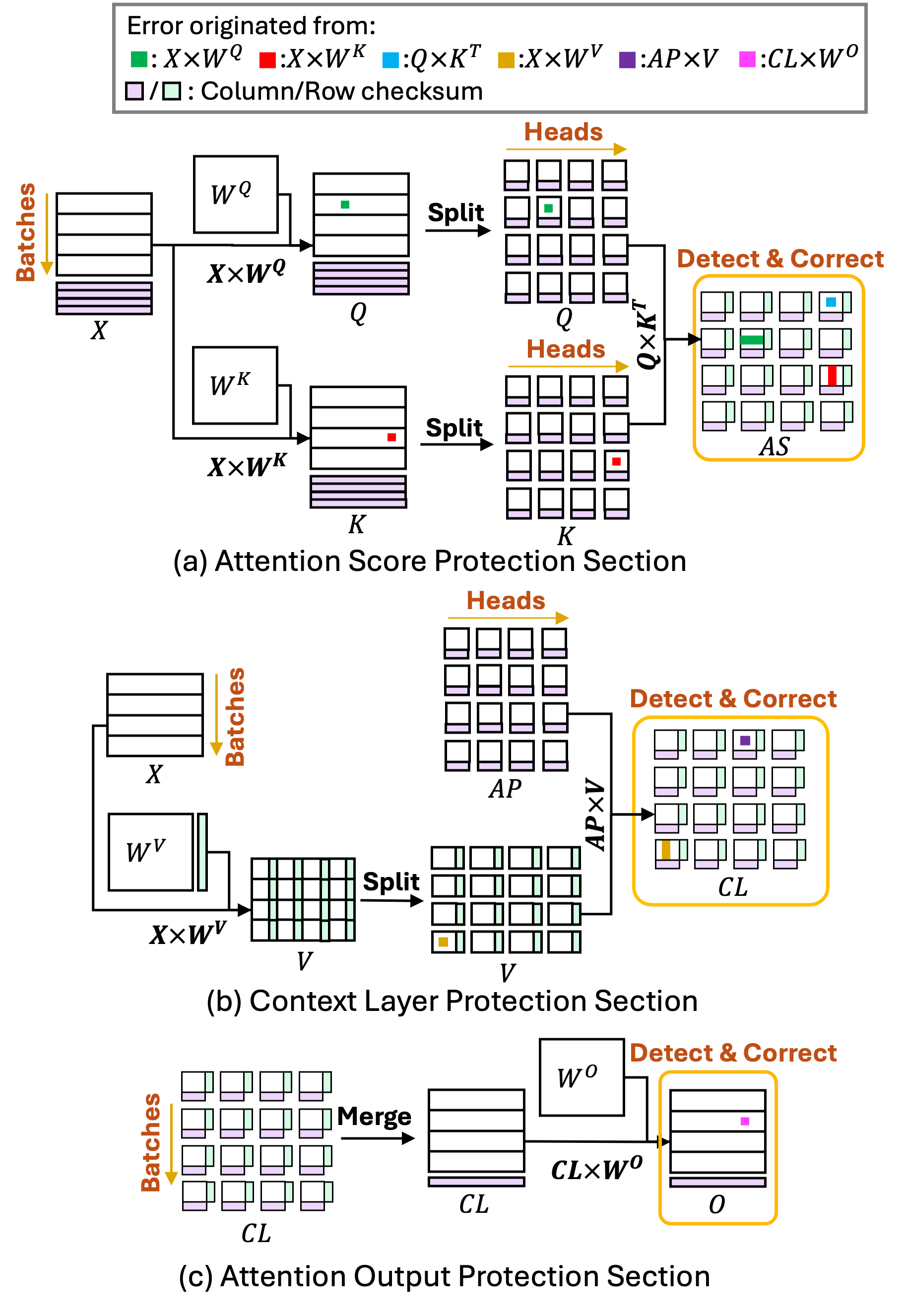}
    \vspace{-2em}
    \caption{Systematic ABFT Schemes in \ours. All 6 \mms Form Three Protection Sections}
    \label{abft-scheme}
    \vspace{-2em}
\end{figure}


\textbf{\textit{Attention Score Protection Section}}: As shown in Figure~\ref{abft-scheme}(a),
to protect the calculation of the attention score, we encode the input matrix $X$ with column checksums. 
After the multiplying input with parameter matrices, the output $Q$ and $K$ matrix will have their corresponding column checksums.
If any computational error occurs during the multiplication, it will manifest as a 0D pattern shown as a red/green element in $Q$ and $K$.
Although we can use their column checksums to correct those errors, we reduce the error handling overhead by delaying detecting and correcting until $Q\times K^T$ is done.
Pre-existing errors in $Q$ or $K$ will propagate as a 1D pattern (shown as a red/green column in $AS$), which is tolerable using \eec.
Finally, 0D errors that occur during $Q\times K^T$ (shown as a blue element in $AS$) can also be corrected immediately after computing $AS$.


\textbf{\textit{Context Layer Protection Section}}: Figure~\ref{abft-scheme}(b) shows our protection mechanism for the calculation of $CL$. We encode the parameter matrix $W^V$ with row checksums. By doing so, $V$ will have row checksums encoded after $X\times W^V$.
Next, we encode $AP$ with column checksum so that after multiplying $AP$ with $V$, the resulting $CL$ will have both row and column checksums. 
If errors occur during $X\times W^V$, it will manifest as a 0D error pattern shown as a yellow element in $V$.
Similarly, we delay the error detection until $CL$ is computed.
Therefore, any 0D error in $V$ will manifest as a 1D error in $CL$, which can be corrected using its checksums.
In addition, 0D errors that occur during $AP\times V$ (shown as purple elements in $CL$) can be corrected using the checksums of $CL$.


\textbf{\textit{Attention Output Protection Section}}: Figure~\ref{abft-scheme}(c) shows the protection on the attention output $O$.
Since it only involves one matrix multiplication, we only need to maintain one side checksum. As we already maintain the column checksums for $CL$ after $AP\times V$, we choose to keep the column checksums and update them along with $CL\times W^O$.
The resulting output matrix $O$ will have column checksums that can detect and correct the 0D errors (shown as magenta elements in $O$) that occurred during $CL\times W^O$.

\subsection{Adaptive ABFT Detection Frequencies}
To make \ours adaptive to operations with different vulnerabilities in \att, we aim to optimize ABFT's detection frequency.
Assuming a system has an error rate (measured by the number of errors per flop that lead to \fffor): $\lambda_{INF}$, $\lambda_{NaN}$, and $\lambda_{nINF}$, we treat the probability distribution of number of error occurrences as the Poisson distribution\cite{chen2023improving}. 
So, the probability of having $k$ errors of type $e$ in an operation $OP$ with total $n_{OP}$ flops can be estimated using $P_{OP}^E(k)=\frac{e^{-\lambda_E\times n_{OP}} (\lambda_E\times n_{OP})^{k}}{k!}$.
Then given protection section $S = \{OP_1, OP_2, ..., OP_m\}$ and a set of error types $E = {INF, NaN, near-INF}$, then the probability of having no error in the section is $R^{free}_{S} = \prod_{i=1}^{m}\prod_{e \in E}^{}P_{OP_i}^{e}(0)$.
Similarly, the probability of having exactly only one error of type $e$ in $OP_j$ and no error for other operations in $S$ is:
\vspace{-0.5em}
$$R^{INF}_{S}(j) = P_{OP_j}^{INF}(1) P_{OP_j}^{NaN}(0) P_{OP_j}^{nINF}(0) R^{free}_{S\smallsetminus OP_j}$$
\vspace{-1em}
$$R^{NaN}_{S}(j) = P_{OP_j}^{INF}(0) P_{OP_j}^{NaN}(1) P_{OP_j}^{nINF}(0) R^{free}_{S\smallsetminus OP_j}$$
\vspace{-1em}
$$R^{nINF}_{S}(j) = P_{OP_j}^{INF}(0) P_{OP_j}^{NaN}(0) P_{OP_j}^{nINF}(1) R^{free}_{S\smallsetminus OP_j}$$

Based on our ABFT scheme design, each segment can tolerate up to one error that occurs in any operation.
As an error does not always lead to non-trainable states, we quantify the vulnerability of an operation by defining $\phi_{OP}^T$ being the probability of a type $e$ error in $OP$ that leads to a non-trainable state, which can be estimated by profiling (Table~\ref{StatTab}).
Assuming $f_{S}$ represents the frequency at which ABFT is enabled for the entire series $S$ (e.g., $f_{S}$ = 0.5 corresponds to enabling ABFT every other time we execute $S$), we define $H^{e}_i = f_{S} + (1-f_{S})\phi_{OP_i}^{e}$. This expression represents the probability of encountering a type $e$ error that can be either correctly handled by ABFT or not handled but does not result in a non-trainable state.
So, we define a quantified estimation of the probability that ABFT can correct all errors in $S$ - fault coverage (FC).
\vspace{-1em}
\begin{equation*}
FC_{S} = R^{free}_{S} + \sum_{i=1}^{m} \sum_{e \in E}^{} R^{e}_S(i)H^{e}_i
\end{equation*}

The FC for the \att can be estimated as:
\vspace{-0.5em}
$$
FC_{att}(f_{AS}, f_{CL}, f_{O}) = FC_{AS}FC_{CL}FC_{O}
$$
To optimize the ABFT frequencies, we define $T_{S}$ as the ABFT overhead the ABFT is added to protect the section $S$. Given a target $FC_{target}$ for \att (e.g., up to 1 error per billion times of execution), finding the optimized frequency can be converted into an optimization problem:

\vspace{-1em}
$$\text{Minimize}\ T = f_{AS}T_{AS} + f_{CL}T_{CL} + f_{O}T_{O}$$
$$ \text{while}\ FC_{att}(f_{AS}, f_{CL}, f_{O}) \geq FC_{target}$$

To solve the above optimization problem, we introduce a term - fault coverage efficiency (FCE) - to estimate how much FC is gained with a time unit of extra ABFT overhead paid for protecting a section $S$:
\vspace{-0.5em}
\begin{equation*}
\begin{split}
& FCE_{S} = \frac{\partial FC_{S}}{\partial T} = \frac{R^{free}_{S} + \sum_{i=1}^{m} \sum_{e \in E}^{} R^{e}_S(i)(1-\phi_{OP_i}^{e})}{T_{S}}
\end{split}
\end{equation*}

Then, finding the optimized ABFT frequencies can be solved using a greedy-based algorithm as shown in Algorithm~\ref{alg-opt-freq}.
The algorithm first computes $FCE$ for all sections and sorts them in descending order.
Then it loops over all sections starting from the most efficient to the least efficient and allocates $t_S$ time for protecting $S$. The maximum time we can allocate is $T_S$. The selection logic is to select as much more efficient protection as possible. Finally, we convert time allocations for each section into protection frequencies.

\SetKwInOut{KwInOut}{In/Out}
\SetKwInOut{KwIn}{In}
\SetKwInOut{KwOut}{Out}
\SetKwFunction{FMain}{OptimizeABFTFrequencies}
\SetKwProg{Fn}{Function}{:}{\KwRet{ $f_{AS}$, $f_{CL}$ ,$f_{O}$}}
\begin{algorithm}[ht!]
\caption{ABFT Frequencies Optimization}
\label{alg-opt-freq}
\Fn{\FMain{$\lambda^{INF}$, $\lambda^{NaN}$, $\lambda^{nINF}$, $\phi$, $FC_{target}$}}{
$FCE_{AS}, FCE_{CL}, FCE_{O} \leftarrow  \lambda^{INF}, \lambda^{NaN}, \lambda^{nINF}, \phi$\\
$FC \leftarrow 0$ ;
$t_{AS}, t_{CL},  t_{O} \leftarrow 0$ \\
\For {$S$ in $DescendSort(FCE)$}{
\If{$FC_{target}-FC < FCE_{S}T_{S}$}{
$t_{S} = T_{S}$ \\
} 
\If{$FC < FC_{target}$}{
$t_{S} = \frac{FC_{target}-FC}{FCE_{S}}$ \\
}
$FC \leftarrow FC + FCE_{S}t_{S}$
}

$f_{AS} \leftarrow \frac{t_{AS}}{T_{AS}}$ ;
$f_{CL} \leftarrow \frac{t_{CL}}{T_{CL}}$ ;
$f_{O} \leftarrow \frac{t_{O}}{T_{O}}$ \\

}
\end{algorithm}
\vspace{-2em}

\subsection{Performance Optimization for GPUs}

\underline{Encoding:} Enabling ABFT for batched matrix operations (e.g., $K\times V$ ) creates an unique computation pattern. This pattern is not well-optimized in vendor-optimized libraries such as NVIDIA cuBLAS. We design highly customized and efficient kernels for encoding checksums to overcome this.
In our kernel, we parallelize the encoding process along the Streaming Multiprocessor by the Number of Heads $\times$ Number of Batches to improve GPU occupancy. To encode each block, we load all data into shared memory and decoupled thread-data mapping between loading and computing~\cite{chen2021accelerating, chen2019tsm2} such that we can achieve fully coalesced global memory accesses while minimizing bank conflict in shared memory.

\noindent
\underline{Updating:} To accelerate the checksum updating operation on GPUs, we pack the checksum with the operand matrix such that the checksum can be updated together with the original operation. We measure the efficiency of updating checksums as fused computation to improve GPU utilization and avoid kernel launch overhead in Section~\ref{sec:abft_overhead}

\noindent
\underline{Detection and Correction: } We also build customized error detection and correction kernels for GPU. As shown in Figure~\ref{PropagationHandling}, error detection and correction are fully parallelizable and divergence-free when no errors occur, introducing minimal overhead to the \att.

\section{Experimental Evaluation}
\subsection{Experiment Methodology}

\textbf{Evaluation Implementation:} We evaluate \ours by integrating it with the latest PyTorch library\cite{paszke2019pytorch}. 
For the best computing efficiency, all core ABFT algorithms are implemented using CUDA and C++ with a few extra control logic added to the Pytorch interface for enabling the functionalities of ABFT in the training process.

\noindent\textbf{Models and Datasets:} To demonstrate the generality of \ours, four LLMs are used in our experiments: Bert\cite{devlin2018bert}, GPT-2\cite{radford2019gpt2}, GPT-Neo\cite{gao2020pile}, and Roberta\cite{liu2019roberta}. To evaluate the training process, we conduct fine-tuning on these models on the Microsoft Research Paraphrase Corpus (MRPC) dataset from the General Language Understanding Evaluation (GLUE) benchmark\cite{wang2018glue} for sequence classification.

\noindent\textbf{Fault Injection:} To estimate how effective \ours is, we inject faults via source code instrumentation. 
We randomly select positions in the output matrix to inject faults to simulate faults that occur during an operation.
INF and NaN are injected via assignments and near-INF is injected by flipping the most significant bit of the selected element.



\noindent\textbf{Experiment Platform: }
Our modified PyTorch is built with CUDA 12.5 and GCC 11. All evaluations are done on a Lambda GPU server equipped with 4 NVIDIA A100 GPUs with 80 GB memory on each GPU and one AMD EPYC 7513 32-core CPU with 1 TB memory. 
We further simulate the performance of training multi-billion parameter LLMs on large-scale multi-GPU systems.

\begin{figure}[t]
    \vspace{-1em}
    \centering
    \includegraphics[width=1\columnwidth]{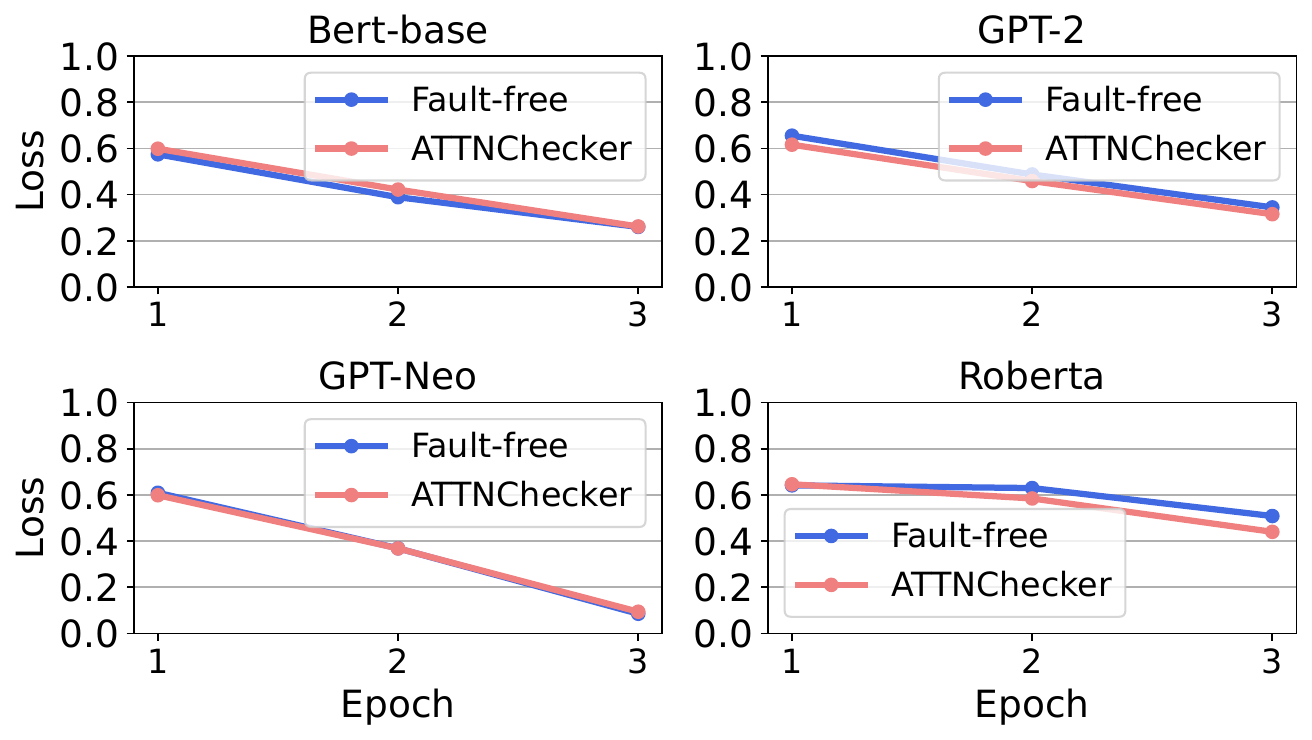}
    \vspace{-2em}
    \caption{Training Loss of Fault-free Execution vs. Faulty Execution Recovered with \ours}
    \label{abft-loss}
\end{figure}
\subsection{Error Detection and Correction Capability}
We first evaluate the effectiveness of \ours by studying its correct rate when different types of faults are injected at different stages of the \att. Each time, we inject one error per execution to a random position of a matrix in the \att with ABFT enabled.
We repeat our fault injection such that we inject a total of 10\% ($\sim$5,000) elements of each output matrix in the \att. Our evaluation of four LLMs shows that \textbf{all errors can be detected and successfully corrected} to their original values. 
Additionally, Figure~\ref{abft-loss} shows that \ours makes a negligible impact on the training loss after error recovery.

\begin{figure}[t]
    \vspace{-1em}
    \centering
    \includegraphics[width=1\columnwidth]{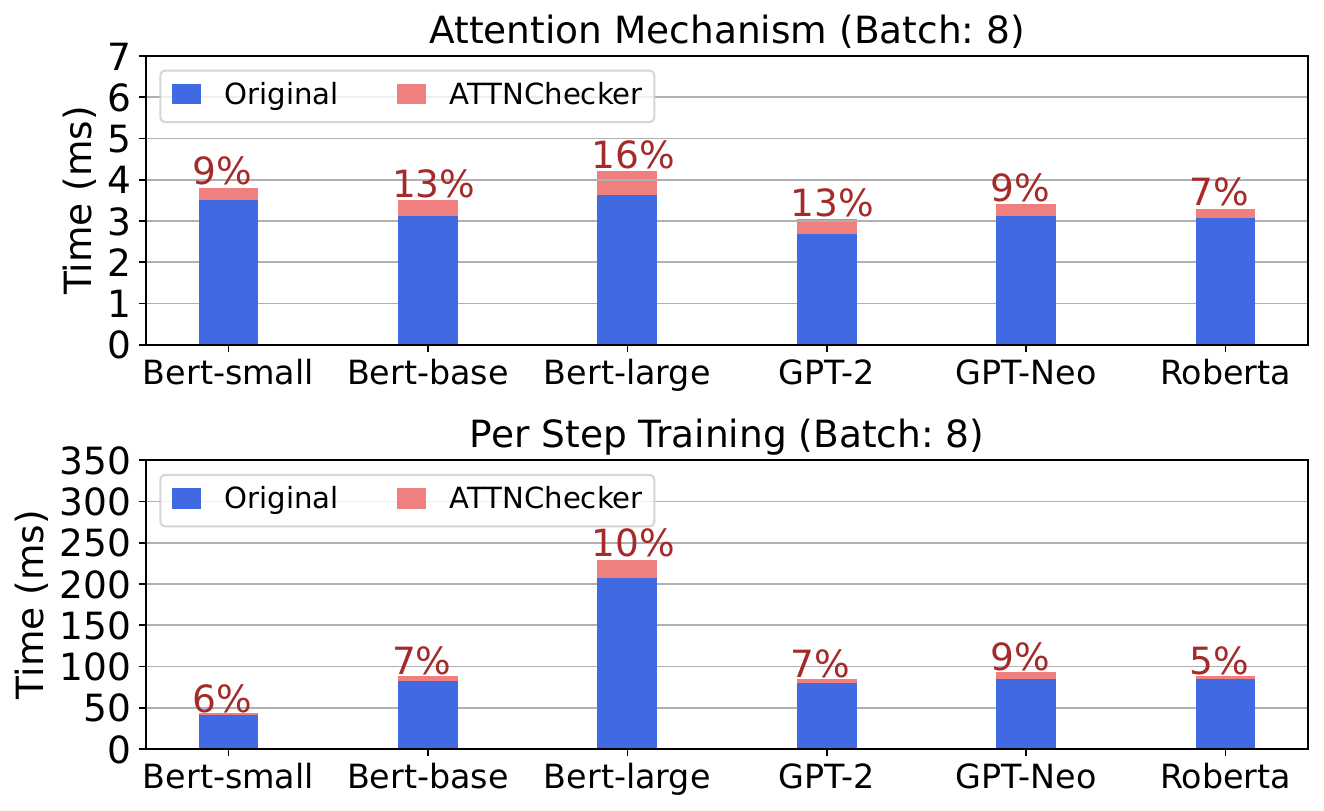}
    \vspace{-2em}
    \caption{\ours Overhead on 6 LLMs}
    \label{model-overhead}
    \vspace{-1.5em}
\end{figure}

\subsection{ABFT Overhead}
\label{sec:abft_overhead}
\begin{figure}[t]
    \centering
    \includegraphics[width=1\columnwidth]{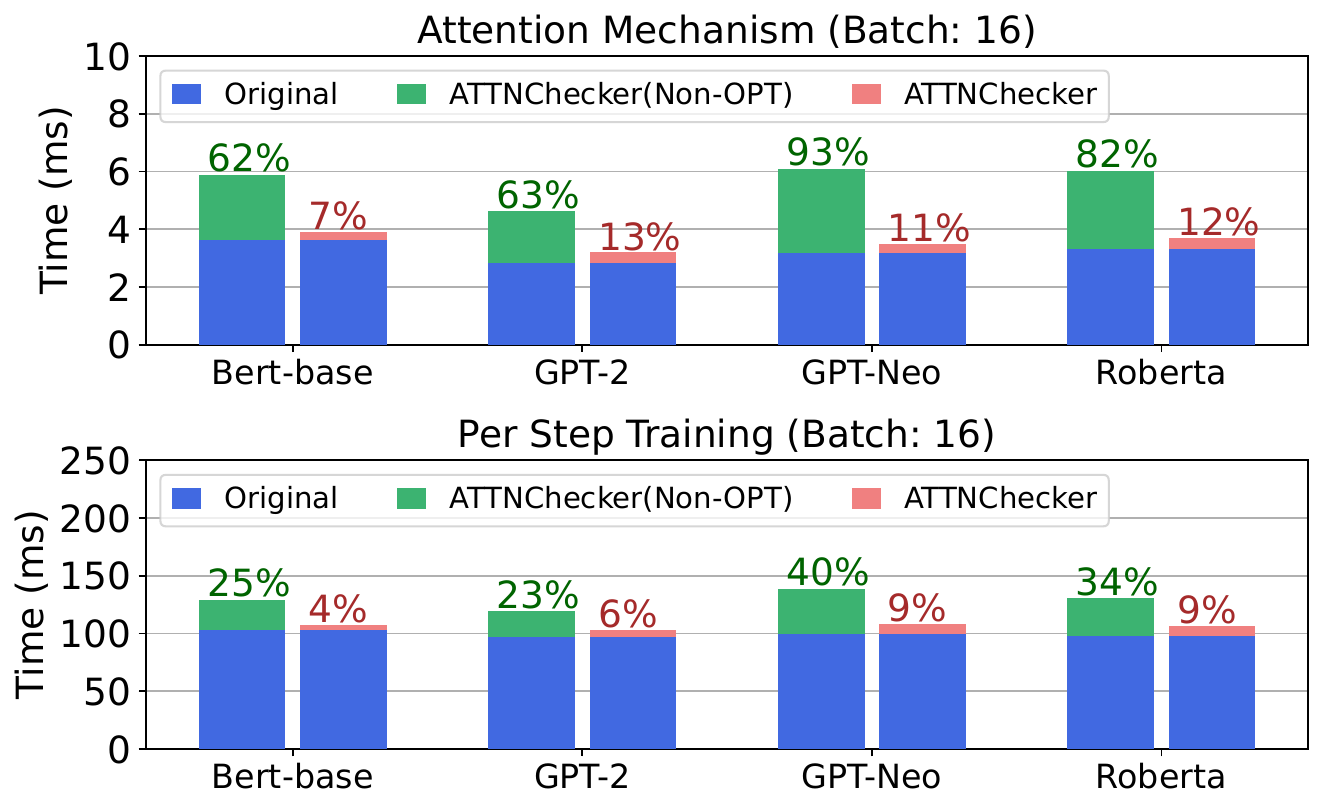}
    \vspace{-2em}
\caption{\ours Overhead With and Without GPU Optimization}
    \label{model-overhead-opt}
    \vspace{-1em}
\end{figure}

\begin{figure}[t]
    \centering
    \includegraphics[width=1\columnwidth]{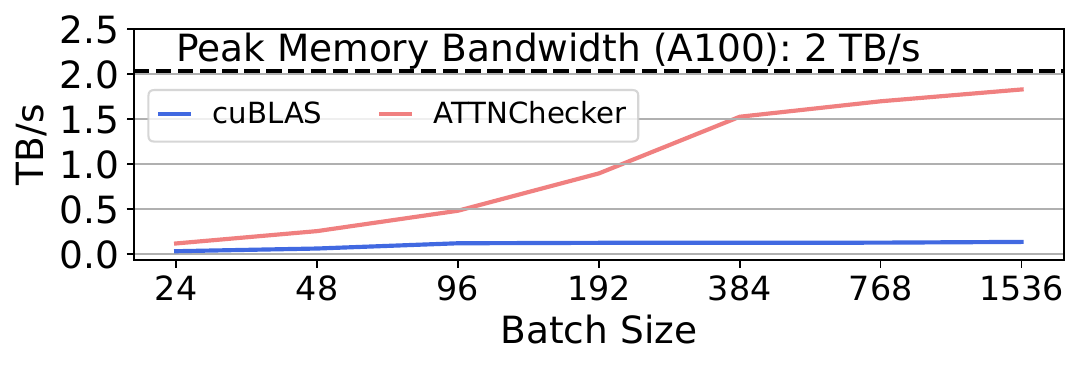}
    \vspace{-2em}
    \caption{Checksum Encoding Throughput}
    \label{chk-enc}
    \vspace{-1em}
\end{figure}

Next, we evaluate the computational overhead caused by \ours for both \att and the end-to-end training process.
Figure~\ref{model-overhead} shows the overhead results across all four models, including three different sizes of the Bert model.
In particular, \ours incurs about $11\%$ overhead to \att block along and $7\%$ overhead for the entire training step on average.
Figure~\ref{model-overhead-opt} shows the overhead of \att for training with and without performance optimization for GPUs.
Our dedicated optimization on GPU reduces ABFT overhead by up to $8.6\times$ for the \att and $6.0\times$ for training.
Furthermore, comparing results of different batch sizes (8 and 16) shows that \ours bring similar overhead across different batch sizes (we only show the result of batch size = 16 due to page limit).
Figure~\ref{chk-enc} shows the checksum encoding throughputs using cuBLAS 12.5 and \ours's optimized encoder kernel. 
Our optimized kernel outperforms cuBLAS by $13\times$ with up to $91.4\%$ memory bandwidth utilization, while cuBLAS only achieves less than $10\%$.

\begin{figure}[t]
    \centering
    \includegraphics[width=1\columnwidth]{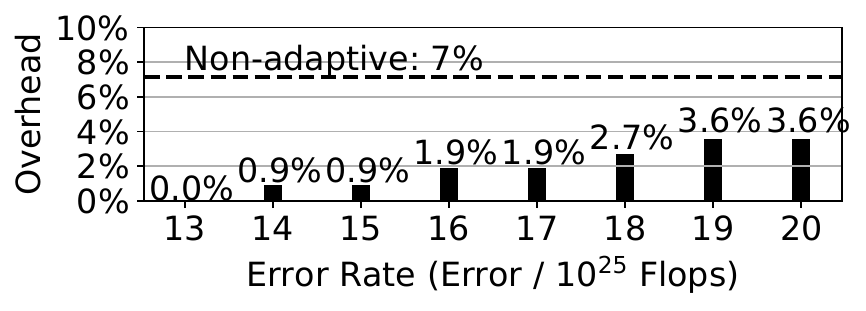}
    \vspace{-2em}
    \caption{Training Overhead Reduction With Optimized ABFT Detection Frequencies}
    \label{overhead-reduction}
    \vspace{-1em}
\end{figure}


\subsection{Adaptive ABFT Detection Frequencies}
To demonstrate how \ours can be adaptive to various error rates representing a wider spectrum of systems, we use our adaptive ABFT optimization algorithm to optimize ABFT detection frequencies for the three detection sections: $S_{AS}$, $S_{CL}$, and $S_{O}$.
We assume the system error rate varies from 13 to 20 errors per $10^{25}$ flops for all three types of error according to the latest field report~\cite{dubey2024llama}. Our target fault coverage is 1 failure per $10^{11}$ executions of \att based on the operation vulnerability of the Bert model.
Figure~\ref{overhead-reduction} shows the per-step training overhead with optimized ABFT checking frequencies. When the error rate is small, we can see that most of the ABFT overhead can be reduced (even to 0 when ABFT is not needed). As the error rate increases, ABFT overhead steadily increases with the detection frequencies yet remains minimal (less than 3.6\%).

\begin{figure}[t]
    \centering
    \includegraphics[width=1\columnwidth]{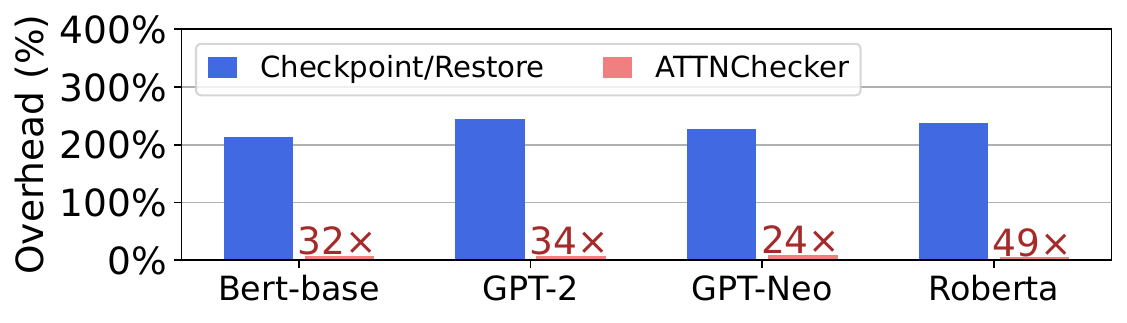}
    \vspace{-2.5em}
    \caption{Per Training Step Recovery Overhead (Checkpoint/Restore vs. \ours)}
    \label{recovery-overhead}
    \vspace{-1em}
\end{figure}

\subsection{Recovery Overhead}
In addition to detection, error correction brings extra overhead. Our evaluation shows that the correcting 1D error caused by error propagation from  $Q$, $K$, and $V$ incurs $0.7\%$ overhead on average. 0D errors can be corrected with 0.3\% overhead on average. Correcting errors in $O$ brings 3.9\% overhead on average since unlike matrices such as $AS$ where error correction is done within a small block, $O$ is a larger matrix merged from smaller blocks, which requires potentially higher recovery cost. 

Figure~\ref{recovery-overhead} shows the pre-training step error recovery overhead of \ours compared with the checkpoint/restore (CR) approach.
Specifically, we assume checkpointing is done per training step and restore is done once the model encounters a non-trainable state.
Since the CR requires loading and re-executing the current training step, the CR causes more than 200\% overhead once a non-trainable state is encountered. 
\ours, on the other hand, for correction-only introduces less than 10\% overhead, which is up to $49\times$ overhead reduction compared with CR.


\subsection{\ours in Scale}
\begin{figure}[t]
    \centering
    \includegraphics[width=1\columnwidth]{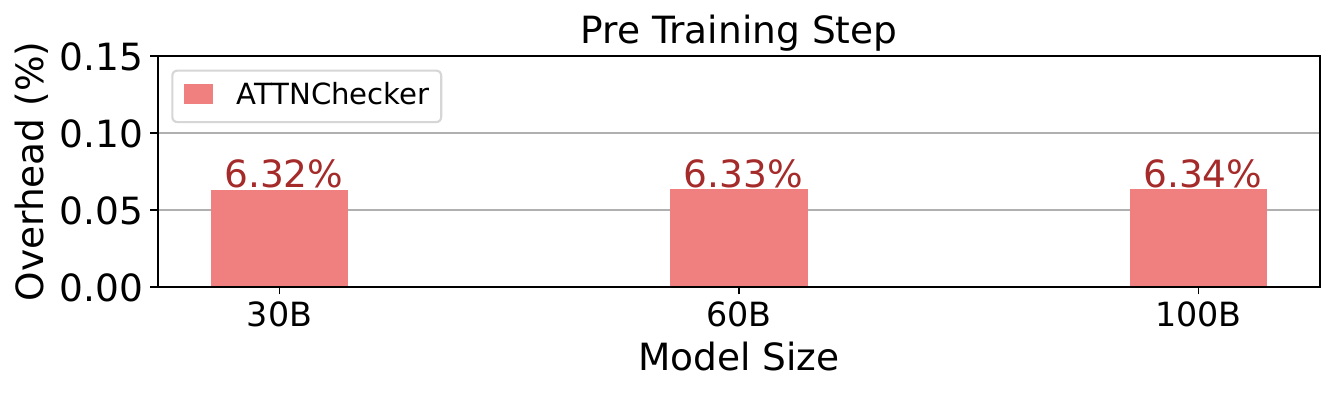}
    \vspace{-2em}
    \caption{ Overhead of \ours with multi-billion parameter LLMs on multi-GPU systems.}
    \vspace{-1em}
    \label{project-perf}
\end{figure}

To further quantify the performance impact of \ours, we simulate the training of multi-billion parameter LLMs on multi-GPU systems using the same simulation methodology as existing work~\cite{lin2024towards}.
Figure \ref{project-perf} presents the overhead of \ours when training LLMs with 30B, 60B, and 100B parameters on 1,024 GPUs using data parallelism. The results demonstrate that the ABFT overhead remains nearly constant as the number of parameters increases, highlighting the scalability and applicability of \ours for large-scale LLM training.

\section{Related Works} 
\label{sec:related}

\textbf{Soft error detection. }
Existing works have explored various techniques to detect and mitigate the impact of soft errors in computing systems. 
For example, some efforts ~\cite{lowcost,relyzer,swift}  investigate instruction-level duplication, which duplicates a subset of critical or error-prone instructions to detect soft errors affecting computations. 
Others~\cite{epvf, trident, fliptracker} prioritize the error-proneness of applications or instructions and select the top ones for duplication or protection, identifying the most critical parts of the computation likely to cause silent data corruptions if affected by soft errors.  
However, these error detection methods are impractical for detecting soft errors in ML models due to their large runtime overhead.

Algorithm-Based Fault Tolerance (ABFT) techniques, on the other hand, are tailored to specific algorithms, leveraging their inherent properties to detect and correct various types of errors, including soft errors, without requiring extensive modifications to the underlying hardware or software. ABFT has been applied to protect and handle errors in matrix decomposition~\cite{huang1984algorithm, wu2014ft, wu2016towards}, and matrix multiplications~\cite{chen2008algorithm, chen2016gpu, chen2018fault, chen2016online}.
As ABFT can effectively protect matrix multiplications, several works have extended it to protect ML models such as CNN~\cite{zhao2020ft} and Deep learning recommendation systems~\cite{li2022efficient}. 
\ours also leverages ABFT but focuses on more computationally intensive and model-quality-critical attention computation in LLMs.

\noindent\textbf{Resilience of ML models under transient faults.} Understanding and evaluating the vulnerability of ML models to hardware faults is becoming increasingly important as ML models scale~\cite{dixit2022detectingsilentdatacorruptions}. 
Existing work investigates various tools and metrics to identify, mitigate, and quantify the impact of faults in ML systems. 
Ares~\cite{8465834} and PyTorchFI~\cite{9151812} are fault injection tools built on PyTorch to help quantify model accuracy drop with faults in ML models.  
Li et al.~\cite{li2017understanding} and Agarwal et al.~\cite{agarwal2023resilience}  investigate the Silent Data Corruption (SDC) rate for CNNs and LLMs, respectively, using fault injection.
While these works provide valuable insights into the resilience of ML models under transient faults, they primarily focus on fault injection and quantifying the impact of faults. In contrast, \ours offers a lightweight, real-time solution for not only detecting but also correcting errors during LLM training. 

\section{Conclusion} \label{sec:conclusion}

The growing complexity and size of LLMs demand effective fault tolerance to maintain efficient and reliable training. 
Traditional fault tolerance mechanisms such as checkpointing are resource-intensive and time-consuming, hindering the efficiency of LLM training.
In response, we introduce \ours, an algorithm-based fault tolerance technique that offers a lightweight, real-time solution for detecting and correcting errors during LLM training.  \ours significantly reduces recovery overheads by up to 49$\times$ compared to traditional methods, while only adding negligible overhead to the training process. By integrating \ours into the PyTorch library and testing on various LLMs, we demonstrate its efficacy in enhancing the resilience and efficiency of LLM training. This advancement provides a robust framework for future fault tolerance in large-scale machine learning models.

\section*{Acknowledgment}

This material is based upon work supported by the U.S. Department of Energy, Office of Science, Office of Advanced Scientific Computing Research, ComPort: Rigorous Testing Methods to Safeguard Software Porting, under Award Number 78284. The platforms used for evaluation in this work are supported by by the U.S. DOE Office of Science, Office of Advanced Scientific Computing Research, under award 66150: "CENATE - Center for Advanced Architecture Evaluation". The Pacific Northwest National Laboratory is operated by Battelle for the U.S. Department of Energy under Contract DE-AC05-76RL01830.

\bibliographystyle{ACM-Reference-Format}
\bibliography{references}


\begin{thebibliography}{47}


\ifx \showCODEN    \undefined \def \showCODEN     #1{\unskip}     \fi
\ifx \showDOI      \undefined \def \showDOI       #1{#1}\fi
\ifx \showISBNx    \undefined \def \showISBNx     #1{\unskip}     \fi
\ifx \showISBNxiii \undefined \def \showISBNxiii  #1{\unskip}     \fi
\ifx \showISSN     \undefined \def \showISSN      #1{\unskip}     \fi
\ifx \showLCCN     \undefined \def \showLCCN      #1{\unskip}     \fi
\ifx \shownote     \undefined \def \shownote      #1{#1}          \fi
\ifx \showarticletitle \undefined \def \showarticletitle #1{#1}   \fi
\ifx \showURL      \undefined \def \showURL       {\relax}        \fi
\providecommand\bibfield[2]{#2}
\providecommand\bibinfo[2]{#2}
\providecommand\natexlab[1]{#1}
\providecommand\showeprint[2][]{arXiv:#2}

\bibitem[Agarwal et~al\mbox{.}(2023)]%
        {agarwal2023resilience}
\bibfield{author}{\bibinfo{person}{Udit~Kumar Agarwal}, \bibinfo{person}{Abraham Chan}, {and} \bibinfo{person}{Karthik Pattabiraman}.} \bibinfo{year}{2023}\natexlab{}.
\newblock \showarticletitle{Resilience Assessment of Large Language Models under Transient Hardware Faults}. In \bibinfo{booktitle}{\emph{2023 IEEE 34th International Symposium on Software Reliability Engineering (ISSRE)}}. IEEE, \bibinfo{pages}{659--670}.
\newblock


\bibitem[Brown(2020)]%
        {brown2020language}
\bibfield{author}{\bibinfo{person}{Tom~B Brown}.} \bibinfo{year}{2020}\natexlab{}.
\newblock \showarticletitle{Language models are few-shot learners}.
\newblock \bibinfo{journal}{\emph{arXiv preprint ArXiv:2005.14165}} (\bibinfo{year}{2020}).
\newblock


\bibitem[Chen et~al\mbox{.}(2018)]%
        {chen2018fault}
\bibfield{author}{\bibinfo{person}{Jieyang Chen}, \bibinfo{person}{Hongbo Li}, \bibinfo{person}{Sihuan Li}, \bibinfo{person}{Xin Liang}, \bibinfo{person}{Panruo Wu}, \bibinfo{person}{Dingwen Tao}, \bibinfo{person}{Kaiming Ouyang}, \bibinfo{person}{Yuanlai Liu}, \bibinfo{person}{Kai Zhao}, \bibinfo{person}{Qiang Guan}, {et~al\mbox{.}}} \bibinfo{year}{2018}\natexlab{}.
\newblock \showarticletitle{Fault tolerant one-sided matrix decompositions on heterogeneous systems with gpus}. In \bibinfo{booktitle}{\emph{SC18: International Conference for High Performance Computing, Networking, Storage and Analysis}}. IEEE, \bibinfo{pages}{854--865}.
\newblock


\bibitem[Chen et~al\mbox{.}(2016a)]%
        {chen2016gpu}
\bibfield{author}{\bibinfo{person}{Jieyang Chen}, \bibinfo{person}{Sihuan Li}, {and} \bibinfo{person}{Zizhong Chen}.} \bibinfo{year}{2016}\natexlab{a}.
\newblock \showarticletitle{Gpu-abft: Optimizing algorithm-based fault tolerance for heterogeneous systems with gpus}. In \bibinfo{booktitle}{\emph{2016 IEEE International Conference on Networking, Architecture and Storage (NAS)}}. IEEE, \bibinfo{pages}{1--2}.
\newblock


\bibitem[Chen et~al\mbox{.}(2016b)]%
        {chen2016online}
\bibfield{author}{\bibinfo{person}{Jieyang Chen}, \bibinfo{person}{Xin Liang}, {and} \bibinfo{person}{Zizhong Chen}.} \bibinfo{year}{2016}\natexlab{b}.
\newblock \showarticletitle{Online algorithm-based fault tolerance for cholesky decomposition on heterogeneous systems with gpus}. In \bibinfo{booktitle}{\emph{2016 IEEE International Parallel and Distributed Processing Symposium (IPDPS)}}. IEEE, \bibinfo{pages}{993--1002}.
\newblock


\bibitem[Chen et~al\mbox{.}(2023)]%
        {chen2023improving}
\bibfield{author}{\bibinfo{person}{Jieyang Chen}, \bibinfo{person}{Xin Liang}, \bibinfo{person}{Kai Zhao}, \bibinfo{person}{Hadi~Zamani Sabzi}, \bibinfo{person}{Laxmi Bhuyan}, {and} \bibinfo{person}{Zizhong Chen}.} \bibinfo{year}{2023}\natexlab{}.
\newblock \showarticletitle{Improving energy saving of one-sided matrix decompositions on cpu-gpu heterogeneous systems}. In \bibinfo{booktitle}{\emph{Proceedings of the 28th ACM SIGPLAN Annual Symposium on Principles and Practice of Parallel Programming}}. \bibinfo{pages}{274--287}.
\newblock


\bibitem[Chen et~al\mbox{.}(2021)]%
        {chen2021accelerating}
\bibfield{author}{\bibinfo{person}{Jieyang Chen}, \bibinfo{person}{Lipeng Wan}, \bibinfo{person}{Xin Liang}, \bibinfo{person}{Ben Whitney}, \bibinfo{person}{Qing Liu}, \bibinfo{person}{David Pugmire}, \bibinfo{person}{Nicholas Thompson}, \bibinfo{person}{Jong~Youl Choi}, \bibinfo{person}{Matthew Wolf}, \bibinfo{person}{Todd Munson}, {et~al\mbox{.}}} \bibinfo{year}{2021}\natexlab{}.
\newblock \showarticletitle{Accelerating multigrid-based hierarchical scientific data refactoring on gpus}. In \bibinfo{booktitle}{\emph{2021 IEEE International Parallel and Distributed Processing Symposium (IPDPS)}}. IEEE, \bibinfo{pages}{859--868}.
\newblock


\bibitem[Chen et~al\mbox{.}(2019)]%
        {chen2019tsm2}
\bibfield{author}{\bibinfo{person}{Jieyang Chen}, \bibinfo{person}{Nan Xiong}, \bibinfo{person}{Xin Liang}, \bibinfo{person}{Dingwen Tao}, \bibinfo{person}{Sihuan Li}, \bibinfo{person}{Kaiming Ouyang}, \bibinfo{person}{Kai Zhao}, \bibinfo{person}{Nathan DeBardeleben}, \bibinfo{person}{Qiang Guan}, {and} \bibinfo{person}{Zizhong Chen}.} \bibinfo{year}{2019}\natexlab{}.
\newblock \showarticletitle{TSM2: optimizing tall-and-skinny matrix-matrix multiplication on GPUs}. In \bibinfo{booktitle}{\emph{Proceedings of the ACM International Conference on Supercomputing}}. \bibinfo{pages}{106--116}.
\newblock


\bibitem[Chen and Dongarra(2008)]%
        {chen2008algorithm}
\bibfield{author}{\bibinfo{person}{Zizhong Chen} {and} \bibinfo{person}{Jack Dongarra}.} \bibinfo{year}{2008}\natexlab{}.
\newblock \showarticletitle{Algorithm-based fault tolerance for fail-stop failures}.
\newblock \bibinfo{journal}{\emph{IEEE Transactions on Parallel and Distributed Systems}} \bibinfo{volume}{19}, \bibinfo{number}{12} (\bibinfo{year}{2008}), \bibinfo{pages}{1628--1641}.
\newblock


\bibitem[Devlin et~al\mbox{.}(2018)]%
        {devlin2018bert}
\bibfield{author}{\bibinfo{person}{Jacob Devlin}, \bibinfo{person}{Ming-Wei Chang}, \bibinfo{person}{Kenton Lee}, {and} \bibinfo{person}{Kristina Toutanova}.} \bibinfo{year}{2018}\natexlab{}.
\newblock \showarticletitle{Bert: Pre-training of deep bidirectional transformers for language understanding}.
\newblock \bibinfo{journal}{\emph{arXiv preprint arXiv:1810.04805}} (\bibinfo{year}{2018}).
\newblock


\bibitem[Dixit et~al\mbox{.}(2022)]%
        {dixit2022detectingsilentdatacorruptions}
\bibfield{author}{\bibinfo{person}{Harish~Dattatraya Dixit}, \bibinfo{person}{Laura Boyle}, \bibinfo{person}{Gautham Vunnam}, \bibinfo{person}{Sneha Pendharkar}, \bibinfo{person}{Matt Beadon}, {and} \bibinfo{person}{Sriram Sankar}.} \bibinfo{year}{2022}\natexlab{}.
\newblock \bibinfo{title}{Detecting silent data corruptions in the wild}.
\newblock
\newblock
\showeprint[arxiv]{2203.08989}~[cs.AR]
\urldef\tempurl%
\url{https://arxiv.org/abs/2203.08989}
\showURL{%
\tempurl}


\bibitem[Dubey et~al\mbox{.}(2024)]%
        {dubey2024llama}
\bibfield{author}{\bibinfo{person}{Abhimanyu Dubey}, \bibinfo{person}{Abhinav Jauhri}, \bibinfo{person}{Abhinav Pandey}, \bibinfo{person}{Abhishek Kadian}, \bibinfo{person}{Ahmad Al-Dahle}, \bibinfo{person}{Aiesha Letman}, \bibinfo{person}{Akhil Mathur}, \bibinfo{person}{Alan Schelten}, \bibinfo{person}{Amy Yang}, \bibinfo{person}{Angela Fan}, {et~al\mbox{.}}} \bibinfo{year}{2024}\natexlab{}.
\newblock \showarticletitle{The llama 3 herd of models}.
\newblock \bibinfo{journal}{\emph{arXiv preprint arXiv:2407.21783}} (\bibinfo{year}{2024}).
\newblock


\bibitem[Fang et~al\mbox{.}(2016)]%
        {epvf}
\bibfield{author}{\bibinfo{person}{B. Fang}, \bibinfo{person}{Q. Lu}, \bibinfo{person}{K. Pattabiraman}, \bibinfo{person}{M. Ripeanu}, {and} \bibinfo{person}{S. Gurumurthi}.} \bibinfo{year}{2016}\natexlab{}.
\newblock \showarticletitle{ePVF: An Enhanced Program Vulnerability Factor Methodology for Cross-Layer Resilience Analysis}. In \bibinfo{booktitle}{\emph{2016 46th Annual IEEE/IFIP International Conference on Dependable Systems and Networks (DSN), acceptance rate = 21\%}}. \bibinfo{pages}{168--179}.
\newblock
\urldef\tempurl%
\url{https://doi.org/10.1109/DSN.2016.24}
\showDOI{\tempurl}


\bibitem[Gao et~al\mbox{.}(2020)]%
        {gao2020pile}
\bibfield{author}{\bibinfo{person}{Leo Gao}, \bibinfo{person}{Stella Biderman}, \bibinfo{person}{Sid Black}, \bibinfo{person}{Laurence Golding}, \bibinfo{person}{Travis Hoppe}, \bibinfo{person}{Charles Foster}, \bibinfo{person}{Jason Phang}, \bibinfo{person}{Horace He}, \bibinfo{person}{Anish Thite}, \bibinfo{person}{Noa Nabeshima}, {et~al\mbox{.}}} \bibinfo{year}{2020}\natexlab{}.
\newblock \showarticletitle{The Pile: An 800GB Dataset of Diverse Text for Language Modeling}.
\newblock \bibinfo{journal}{\emph{arXiv preprint arXiv:2101.00027}} (\bibinfo{year}{2020}).
\newblock


\bibitem[Gonçalves~de Oliveira et~al\mbox{.}(2016)]%
        {raditation-induced}
\bibfield{author}{\bibinfo{person}{Daniel Alfonso~Gonçalves Gonçalves~de Oliveira}, \bibinfo{person}{Laercio~Lima Pilla}, \bibinfo{person}{Thiago Santini}, {and} \bibinfo{person}{Paolo Rech}.} \bibinfo{year}{2016}\natexlab{}.
\newblock \showarticletitle{Evaluation and Mitigation of Radiation-Induced Soft Errors in Graphics Processing Units}.
\newblock \bibinfo{journal}{\emph{IEEE Trans. Comput.}} \bibinfo{volume}{65}, \bibinfo{number}{3} (\bibinfo{year}{2016}), \bibinfo{pages}{791--804}.
\newblock
\urldef\tempurl%
\url{https://doi.org/10.1109/TC.2015.2444855}
\showDOI{\tempurl}


\bibitem[Guo et~al\mbox{.}(2018)]%
        {fliptracker}
\bibfield{author}{\bibinfo{person}{Luanzheng Guo}, \bibinfo{person}{Dong Li}, \bibinfo{person}{Ignacio Laguna}, {and} \bibinfo{person}{Martin Schulz}.} \bibinfo{year}{2018}\natexlab{}.
\newblock \showarticletitle{FlipTracker: Understanding Natural Error Resilience in HPC Applications}. In \bibinfo{booktitle}{\emph{SC18: International Conference for High Performance Computing, Networking, Storage and Analysis}}. \bibinfo{pages}{94--107}.
\newblock
\urldef\tempurl%
\url{https://doi.org/10.1109/SC.2018.00011}
\showDOI{\tempurl}


\bibitem[Hari et~al\mbox{.}(2012a)]%
        {lowcost}
\bibfield{author}{\bibinfo{person}{Siva Kumar~Sastry Hari}, \bibinfo{person}{Sarita~V Adve}, {and} \bibinfo{person}{Helia Naeimi}.} \bibinfo{year}{2012}\natexlab{a}.
\newblock \showarticletitle{{Low-cost program-level detectors for reducing silent data corruptions}}. In \bibinfo{booktitle}{\emph{2012 42nd Annual IEEE/IFIP International Conference on Dependable Systems and Networks (DSN)}}. \bibinfo{publisher}{IEEE}, \bibinfo{pages}{1--12}.
\newblock


\bibitem[Hari et~al\mbox{.}(2012b)]%
        {relyzer}
\bibfield{author}{\bibinfo{person}{Siva Kumar~Sastry Hari}, \bibinfo{person}{Sarita~V. Adve}, \bibinfo{person}{Helia Naeimi}, {and} \bibinfo{person}{Pradeep Ramachandran}.} \bibinfo{year}{2012}\natexlab{b}.
\newblock \showarticletitle{{Relyzer}: exploiting application-level fault equivalence to analyze application resiliency to transient faults}. In \bibinfo{booktitle}{\emph{ACM International Conference on Architectural Support for Programming Languages and Operating Systems}}.
\newblock


\bibitem[He et~al\mbox{.}(2023)]%
        {he2023understanding}
\bibfield{author}{\bibinfo{person}{Yi He}, \bibinfo{person}{Mike Hutton}, \bibinfo{person}{Steven Chan}, \bibinfo{person}{Robert De~Gruijl}, \bibinfo{person}{Rama Govindaraju}, \bibinfo{person}{Nishant Patil}, {and} \bibinfo{person}{Yanjing Li}.} \bibinfo{year}{2023}\natexlab{}.
\newblock \showarticletitle{Understanding and mitigating hardware failures in deep learning training systems}. In \bibinfo{booktitle}{\emph{Proceedings of the 50th Annual International Symposium on Computer Architecture}}. \bibinfo{pages}{1--16}.
\newblock


\bibitem[Hoffmann et~al\mbox{.}(2022)]%
        {hoffmann2022training}
\bibfield{author}{\bibinfo{person}{Jordan Hoffmann}, \bibinfo{person}{Sebastian Borgeaud}, \bibinfo{person}{Arthur Mensch}, \bibinfo{person}{Elena Buchatskaya}, \bibinfo{person}{Trevor Cai}, \bibinfo{person}{Eliza Rutherford}, \bibinfo{person}{Diego de~Las Casas}, \bibinfo{person}{Lisa~Anne Hendricks}, \bibinfo{person}{Johannes Welbl}, \bibinfo{person}{Aidan Clark}, {et~al\mbox{.}}} \bibinfo{year}{2022}\natexlab{}.
\newblock \showarticletitle{Training compute-optimal large language models}.
\newblock \bibinfo{journal}{\emph{arXiv preprint arXiv:2203.15556}} (\bibinfo{year}{2022}).
\newblock


\bibitem[Huang and Abraham(1984)]%
        {huang1984algorithm}
\bibfield{author}{\bibinfo{person}{Kuang-Hua Huang} {and} \bibinfo{person}{Jacob~A Abraham}.} \bibinfo{year}{1984}\natexlab{}.
\newblock \showarticletitle{Algorithm-based fault tolerance for matrix operations}.
\newblock \bibinfo{journal}{\emph{IEEE transactions on computers}} \bibinfo{volume}{100}, \bibinfo{number}{6} (\bibinfo{year}{1984}), \bibinfo{pages}{518--528}.
\newblock


\bibitem[Jiang et~al\mbox{.}(2024)]%
        {jiang2024megascale}
\bibfield{author}{\bibinfo{person}{Ziheng Jiang}, \bibinfo{person}{Haibin Lin}, \bibinfo{person}{Yinmin Zhong}, \bibinfo{person}{Qi Huang}, \bibinfo{person}{Yangrui Chen}, \bibinfo{person}{Zhi Zhang}, \bibinfo{person}{Yanghua Peng}, \bibinfo{person}{Xiang Li}, \bibinfo{person}{Cong Xie}, \bibinfo{person}{Shibiao Nong}, {et~al\mbox{.}}} \bibinfo{year}{2024}\natexlab{}.
\newblock \showarticletitle{$\{$MegaScale$\}$: Scaling large language model training to more than 10,000 $\{$GPUs$\}$}. In \bibinfo{booktitle}{\emph{21st USENIX Symposium on Networked Systems Design and Implementation (NSDI 24)}}. \bibinfo{pages}{745--760}.
\newblock


\bibitem[Li et~al\mbox{.}(2017)]%
        {li2017understanding}
\bibfield{author}{\bibinfo{person}{Guanpeng Li}, \bibinfo{person}{Siva Kumar~Sastry Hari}, \bibinfo{person}{Michael Sullivan}, \bibinfo{person}{Timothy Tsai}, \bibinfo{person}{Karthik Pattabiraman}, \bibinfo{person}{Joel Emer}, {and} \bibinfo{person}{Stephen~W Keckler}.} \bibinfo{year}{2017}\natexlab{}.
\newblock \showarticletitle{Understanding error propagation in deep learning neural network (DNN) accelerators and applications}. In \bibinfo{booktitle}{\emph{Proceedings of the International Conference for High Performance Computing, Networking, Storage and Analysis}}. \bibinfo{pages}{1--12}.
\newblock


\bibitem[{Li} et~al\mbox{.}(2018)]%
        {trident}
\bibfield{author}{\bibinfo{person}{G. {Li}}, \bibinfo{person}{K. {Pattabiraman}}, \bibinfo{person}{S.~K.~S. {Hari}}, \bibinfo{person}{M. {Sullivan}}, {and} \bibinfo{person}{T. {Tsai}}.} \bibinfo{year}{2018}\natexlab{}.
\newblock \showarticletitle{Modeling Soft-Error Propagation in Programs}. In \bibinfo{booktitle}{\emph{2018 48th Annual IEEE/IFIP International Conference on Dependable Systems and Networks (DSN)}}. \bibinfo{pages}{27--38}.
\newblock
\urldef\tempurl%
\url{https://doi.org/10.1109/DSN.2018.00016}
\showDOI{\tempurl}


\bibitem[Li et~al\mbox{.}(2022)]%
        {li2022efficient}
\bibfield{author}{\bibinfo{person}{Sihuan Li}, \bibinfo{person}{Jianyu Huang}, \bibinfo{person}{Ping Tak~Peter Tang}, \bibinfo{person}{Daya Khudia}, \bibinfo{person}{Jongsoo Park}, \bibinfo{person}{Harish~Dattatraya Dixit}, {and} \bibinfo{person}{Zizhong Chen}.} \bibinfo{year}{2022}\natexlab{}.
\newblock \showarticletitle{Efficient soft-error detection for low-precision deep learning recommendation models}. In \bibinfo{booktitle}{\emph{2022 IEEE International Conference on Big Data (Big Data)}}. IEEE, \bibinfo{pages}{1556--1563}.
\newblock


\bibitem[Liang et~al\mbox{.}(2017)]%
        {liang2017correcting}
\bibfield{author}{\bibinfo{person}{Xin Liang}, \bibinfo{person}{Jieyang Chen}, \bibinfo{person}{Dingwen Tao}, \bibinfo{person}{Sihuan Li}, \bibinfo{person}{Panruo Wu}, \bibinfo{person}{Hongbo Li}, \bibinfo{person}{Kaiming Ouyang}, \bibinfo{person}{Yuanlai Liu}, \bibinfo{person}{Fengguang Song}, {and} \bibinfo{person}{Zizhong Chen}.} \bibinfo{year}{2017}\natexlab{}.
\newblock \showarticletitle{Correcting soft errors online in fast fourier transform}. In \bibinfo{booktitle}{\emph{Proceedings of the International Conference for High Performance Computing, Networking, Storage and Analysis}}. \bibinfo{pages}{1--12}.
\newblock


\bibitem[Lin et~al\mbox{.}(2024)]%
        {lin2024towards}
\bibfield{author}{\bibinfo{person}{Zhongyi Lin}, \bibinfo{person}{Ning Sun}, \bibinfo{person}{Pallab Bhattacharya}, \bibinfo{person}{Xizhou Feng}, \bibinfo{person}{Louis Feng}, {and} \bibinfo{person}{John~D Owens}.} \bibinfo{year}{2024}\natexlab{}.
\newblock \showarticletitle{Towards Universal Performance Modeling for Machine Learning Training on Multi-GPU Platforms}.
\newblock \bibinfo{journal}{\emph{arXiv preprint arXiv:2404.12674}} (\bibinfo{year}{2024}).
\newblock


\bibitem[Liu et~al\mbox{.}(2019)]%
        {liu2019roberta}
\bibfield{author}{\bibinfo{person}{Yinhan Liu}, \bibinfo{person}{Myle Ott}, \bibinfo{person}{Naman Goyal}, \bibinfo{person}{Jingfei Du}, \bibinfo{person}{Mandar Joshi}, \bibinfo{person}{Danqi Chen}, \bibinfo{person}{Omer Levy}, \bibinfo{person}{Mike Lewis}, \bibinfo{person}{Luke Zettlemoyer}, {and} \bibinfo{person}{Veselin Stoyanov}.} \bibinfo{year}{2019}\natexlab{}.
\newblock \showarticletitle{Roberta: A robustly optimized bert pretraining approach}.
\newblock \bibinfo{journal}{\emph{arXiv preprint arXiv:1907.11692}} (\bibinfo{year}{2019}).
\newblock


\bibitem[Mahmoud et~al\mbox{.}(2020)]%
        {9151812}
\bibfield{author}{\bibinfo{person}{Abdulrahman Mahmoud}, \bibinfo{person}{Neeraj Aggarwal}, \bibinfo{person}{Alex Nobbe}, \bibinfo{person}{Jose Rodrigo~Sanchez Vicarte}, \bibinfo{person}{Sarita~V. Adve}, \bibinfo{person}{Christopher~W. Fletcher}, \bibinfo{person}{Iuri Frosio}, {and} \bibinfo{person}{Siva Kumar~Sastry Hari}.} \bibinfo{year}{2020}\natexlab{}.
\newblock \showarticletitle{PyTorchFI: A Runtime Perturbation Tool for DNNs}. In \bibinfo{booktitle}{\emph{2020 50th Annual IEEE/IFIP International Conference on Dependable Systems and Networks Workshops (DSN-W)}}. \bibinfo{pages}{25--31}.
\newblock
\urldef\tempurl%
\url{https://doi.org/10.1109/DSN-W50199.2020.00014}
\showDOI{\tempurl}


\bibitem[Martino et~al\mbox{.}(2015)]%
        {extremescaleresilience}
\bibfield{author}{\bibinfo{person}{{Catello Di} Martino}, \bibinfo{person}{William Kramer}, \bibinfo{person}{Zbigniew Kalbarczyk}, {and} \bibinfo{person}{Ravishankar Iyer}.} \bibinfo{year}{2015}\natexlab{}.
\newblock \showarticletitle{Measuring and Understanding Extreme-Scale Application Resilience: A Field Study of 5,000,000 HPC Application Runs} \emph{(\bibinfo{series}{Proceedings of the International Conference on Dependable Systems and Networks})}. \bibinfo{publisher}{IEEE Computer Society}, \bibinfo{pages}{25--36}.
\newblock
\urldef\tempurl%
\url{https://doi.org/10.1109/DSN.2015.50}
\showDOI{\tempurl}


\bibitem[Nie et~al\mbox{.}(2016)]%
        {largescale_gpu}
\bibfield{author}{\bibinfo{person}{Bin Nie}, \bibinfo{person}{Devesh Tiwari}, \bibinfo{person}{Saurabh Gupta}, \bibinfo{person}{Evgenia Smirni}, {and} \bibinfo{person}{James~H. Rogers}.} \bibinfo{year}{2016}\natexlab{}.
\newblock \showarticletitle{A large-scale study of soft-errors on GPUs in the field}. In \bibinfo{booktitle}{\emph{2016 IEEE International Symposium on High Performance Computer Architecture (HPCA)}}. \bibinfo{pages}{519--530}.
\newblock
\urldef\tempurl%
\url{https://doi.org/10.1109/HPCA.2016.7446091}
\showDOI{\tempurl}


\bibitem[Paszke et~al\mbox{.}(2019)]%
        {paszke2019pytorch}
\bibfield{author}{\bibinfo{person}{Adam Paszke}, \bibinfo{person}{Sam Gross}, \bibinfo{person}{Francisco Massa}, \bibinfo{person}{Adam Lerer}, \bibinfo{person}{James Bradbury}, \bibinfo{person}{Gregory Chanan}, \bibinfo{person}{Trevor Killeen}, \bibinfo{person}{Zeming Lin}, \bibinfo{person}{Natalia Gimelshein}, \bibinfo{person}{Luca Antiga}, {et~al\mbox{.}}} \bibinfo{year}{2019}\natexlab{}.
\newblock \showarticletitle{Pytorch: An imperative style, high-performance deep learning library}.
\newblock \bibinfo{journal}{\emph{Advances in neural information processing systems}}  \bibinfo{volume}{32} (\bibinfo{year}{2019}).
\newblock


\bibitem[Radford et~al\mbox{.}(2019)]%
        {radford2019gpt2}
\bibfield{author}{\bibinfo{person}{Alec Radford}, \bibinfo{person}{Jeffrey Wu}, \bibinfo{person}{Rewon Child}, \bibinfo{person}{David Luan}, \bibinfo{person}{Dario Amodei}, \bibinfo{person}{Ilya Sutskever}, {et~al\mbox{.}}} \bibinfo{year}{2019}\natexlab{}.
\newblock \showarticletitle{Language models are unsupervised multitask learners}.
\newblock \bibinfo{journal}{\emph{OpenAI blog}} \bibinfo{volume}{1}, \bibinfo{number}{8} (\bibinfo{year}{2019}), \bibinfo{pages}{9}.
\newblock


\bibitem[Rajbhandari et~al\mbox{.}(2020)]%
        {rajbhandari2020zero}
\bibfield{author}{\bibinfo{person}{Samyam Rajbhandari}, \bibinfo{person}{Jeff Rasley}, \bibinfo{person}{Olatunji Ruwase}, {and} \bibinfo{person}{Yuxiong He}.} \bibinfo{year}{2020}\natexlab{}.
\newblock \showarticletitle{Zero: Memory optimizations toward training trillion parameter models}. In \bibinfo{booktitle}{\emph{SC20: International Conference for High Performance Computing, Networking, Storage and Analysis}}. IEEE, \bibinfo{pages}{1--16}.
\newblock


\bibitem[Reagen et~al\mbox{.}(2018)]%
        {8465834}
\bibfield{author}{\bibinfo{person}{Brandon Reagen}, \bibinfo{person}{Udit Gupta}, \bibinfo{person}{Lillian Pentecost}, \bibinfo{person}{Paul Whatmough}, \bibinfo{person}{Sae~Kyu Lee}, \bibinfo{person}{Niamh Mulholland}, \bibinfo{person}{David Brooks}, {and} \bibinfo{person}{Gu-Yeon Wei}.} \bibinfo{year}{2018}\natexlab{}.
\newblock \showarticletitle{Ares: A framework for quantifying the resilience of deep neural networks}. In \bibinfo{booktitle}{\emph{2018 55th ACM/ESDA/IEEE Design Automation Conference (DAC)}}. \bibinfo{pages}{1--6}.
\newblock
\urldef\tempurl%
\url{https://doi.org/10.1109/DAC.2018.8465834}
\showDOI{\tempurl}


\bibitem[Reis et~al\mbox{.}(2005)]%
        {swift}
\bibfield{author}{\bibinfo{person}{G~A Reis}, \bibinfo{person}{J Chang}, \bibinfo{person}{N Vachharajani}, \bibinfo{person}{R Rangan}, {and} \bibinfo{person}{D~I August}.} \bibinfo{year}{2005}\natexlab{}.
\newblock \showarticletitle{{SWIFT: Software Implemented Fault Tolerance}}. In \bibinfo{booktitle}{\emph{International Symposium on Code Generation and Optimization}}. \bibinfo{publisher}{IEEE}, \bibinfo{pages}{243--254}.
\newblock


\bibitem[Vaswani et~al\mbox{.}(2017)]%
        {vaswani2017attention}
\bibfield{author}{\bibinfo{person}{Ashish Vaswani}, \bibinfo{person}{Noam Shazeer}, \bibinfo{person}{Niki Parmar}, \bibinfo{person}{Jakob Uszkoreit}, \bibinfo{person}{Llion Jones}, \bibinfo{person}{Aidan~N Gomez}, \bibinfo{person}{{\L}ukasz Kaiser}, {and} \bibinfo{person}{Illia Polosukhin}.} \bibinfo{year}{2017}\natexlab{}.
\newblock \showarticletitle{Attention is all you need}.
\newblock \bibinfo{journal}{\emph{Advances in neural information processing systems}}  \bibinfo{volume}{30} (\bibinfo{year}{2017}).
\newblock


\bibitem[Wan et~al\mbox{.}(2024)]%
        {wan2024bytecheckpoint}
\bibfield{author}{\bibinfo{person}{Borui Wan}, \bibinfo{person}{Mingji Han}, \bibinfo{person}{Yiyao Sheng}, \bibinfo{person}{Zhichao Lai}, \bibinfo{person}{Mofan Zhang}, \bibinfo{person}{Junda Zhang}, \bibinfo{person}{Yanghua Peng}, \bibinfo{person}{Haibin Lin}, \bibinfo{person}{Xin Liu}, {and} \bibinfo{person}{Chuan Wu}.} \bibinfo{year}{2024}\natexlab{}.
\newblock \showarticletitle{ByteCheckpoint: A Unified Checkpointing System for LLM Development}.
\newblock \bibinfo{journal}{\emph{arXiv preprint arXiv:2407.20143}} (\bibinfo{year}{2024}).
\newblock


\bibitem[Wang et~al\mbox{.}(2018)]%
        {wang2018glue}
\bibfield{author}{\bibinfo{person}{Alex Wang}, \bibinfo{person}{Amanpreet Singh}, \bibinfo{person}{Julian Michael}, \bibinfo{person}{Felix Hill}, \bibinfo{person}{Omer Levy}, {and} \bibinfo{person}{Samuel~R Bowman}.} \bibinfo{year}{2018}\natexlab{}.
\newblock \showarticletitle{GLUE: A multi-task benchmark and analysis platform for natural language understanding}.
\newblock \bibinfo{journal}{\emph{arXiv preprint arXiv:1804.07461}} (\bibinfo{year}{2018}).
\newblock


\bibitem[Wang et~al\mbox{.}(2023)]%
        {wang2023reliable}
\bibfield{author}{\bibinfo{person}{Yuxin Wang}, \bibinfo{person}{Shaohuai Shi}, \bibinfo{person}{Xin He}, \bibinfo{person}{Zhenheng Tang}, \bibinfo{person}{Xinglin Pan}, \bibinfo{person}{Yang Zheng}, \bibinfo{person}{Xiaoyu Wu}, \bibinfo{person}{Amelie~Chi Zhou}, \bibinfo{person}{Bingsheng He}, {and} \bibinfo{person}{Xiaowen Chu}.} \bibinfo{year}{2023}\natexlab{}.
\newblock \showarticletitle{Reliable and Efficient In-Memory Fault Tolerance of Large Language Model Pretraining}.
\newblock \bibinfo{journal}{\emph{arXiv preprint arXiv:2310.12670}} (\bibinfo{year}{2023}).
\newblock


\bibitem[Wu et~al\mbox{.}(2023)]%
        {wu2023transom}
\bibfield{author}{\bibinfo{person}{Baodong Wu}, \bibinfo{person}{Lei Xia}, \bibinfo{person}{Qingping Li}, \bibinfo{person}{Kangyu Li}, \bibinfo{person}{Xu Chen}, \bibinfo{person}{Yongqiang Guo}, \bibinfo{person}{Tieyao Xiang}, \bibinfo{person}{Yuheng Chen}, {and} \bibinfo{person}{Shigang Li}.} \bibinfo{year}{2023}\natexlab{}.
\newblock \showarticletitle{Transom: An efficient fault-tolerant system for training llms}.
\newblock \bibinfo{journal}{\emph{arXiv preprint arXiv:2310.10046}} (\bibinfo{year}{2023}).
\newblock


\bibitem[Wu and Chen(2014)]%
        {wu2014ft}
\bibfield{author}{\bibinfo{person}{Panruo Wu} {and} \bibinfo{person}{Zizhong Chen}.} \bibinfo{year}{2014}\natexlab{}.
\newblock \showarticletitle{FT-ScaLAPACK: Correcting soft errors on-line for ScaLAPACK Cholesky, QR, and LU factorization routines}. In \bibinfo{booktitle}{\emph{Proceedings of the 23rd international symposium on High-performance parallel and distributed computing}}. \bibinfo{pages}{49--60}.
\newblock


\bibitem[Wu et~al\mbox{.}(2017)]%
        {wu2017silent}
\bibfield{author}{\bibinfo{person}{Panruo Wu}, \bibinfo{person}{Nathan DeBardeleben}, \bibinfo{person}{Qiang Guan}, \bibinfo{person}{Sean Blanchard}, \bibinfo{person}{Jieyang Chen}, \bibinfo{person}{Dingwen Tao}, \bibinfo{person}{Xin Liang}, \bibinfo{person}{Kaiming Ouyang}, {and} \bibinfo{person}{Zizhong Chen}.} \bibinfo{year}{2017}\natexlab{}.
\newblock \showarticletitle{Silent data corruption resilient two-sided matrix factorizations}. In \bibinfo{booktitle}{\emph{Proceedings of the 22nd ACM SIGPLAN Symposium on Principles and Practice of Parallel Programming}}. \bibinfo{pages}{415--427}.
\newblock


\bibitem[Wu et~al\mbox{.}(2016)]%
        {wu2016towards}
\bibfield{author}{\bibinfo{person}{Panruo Wu}, \bibinfo{person}{Qiang Guan}, \bibinfo{person}{Nathan DeBardeleben}, \bibinfo{person}{Sean Blanchard}, \bibinfo{person}{Dingwen Tao}, \bibinfo{person}{Xin Liang}, \bibinfo{person}{Jieyang Chen}, {and} \bibinfo{person}{Zizhong Chen}.} \bibinfo{year}{2016}\natexlab{}.
\newblock \showarticletitle{Towards practical algorithm based fault tolerance in dense linear algebra}. In \bibinfo{booktitle}{\emph{Proceedings of the 25th ACM International Symposium on High-Performance Parallel and Distributed Computing}}. \bibinfo{pages}{31--42}.
\newblock


\bibitem[Xu et~al\mbox{.}(2021)]%
        {xu2021learning}
\bibfield{author}{\bibinfo{person}{Chunmei Xu}, \bibinfo{person}{Shengheng Liu}, \bibinfo{person}{Zhaohui Yang}, \bibinfo{person}{Yongming Huang}, {and} \bibinfo{person}{Kai-Kit Wong}.} \bibinfo{year}{2021}\natexlab{}.
\newblock \showarticletitle{Learning rate optimization for federated learning exploiting over-the-air computation}.
\newblock \bibinfo{journal}{\emph{IEEE Journal on Selected Areas in Communications}} \bibinfo{volume}{39}, \bibinfo{number}{12} (\bibinfo{year}{2021}), \bibinfo{pages}{3742--3756}.
\newblock


\bibitem[Zhang et~al\mbox{.}(2019)]%
        {zhang2019quantifying}
\bibfield{author}{\bibinfo{person}{Zhao Zhang}, \bibinfo{person}{Lei Huang}, \bibinfo{person}{Ruizhu Huang}, \bibinfo{person}{Weijia Xu}, {and} \bibinfo{person}{Daniel~S Katz}.} \bibinfo{year}{2019}\natexlab{}.
\newblock \showarticletitle{Quantifying the impact of memory errors in deep learning}. In \bibinfo{booktitle}{\emph{2019 IEEE International Conference on Cluster Computing (CLUSTER)}}. IEEE, \bibinfo{pages}{1--12}.
\newblock


\bibitem[Zhao et~al\mbox{.}(2020)]%
        {zhao2020ft}
\bibfield{author}{\bibinfo{person}{Kai Zhao}, \bibinfo{person}{Sheng Di}, \bibinfo{person}{Sihuan Li}, \bibinfo{person}{Xin Liang}, \bibinfo{person}{Yujia Zhai}, \bibinfo{person}{Jieyang Chen}, \bibinfo{person}{Kaiming Ouyang}, \bibinfo{person}{Franck Cappello}, {and} \bibinfo{person}{Zizhong Chen}.} \bibinfo{year}{2020}\natexlab{}.
\newblock \showarticletitle{FT-CNN: Algorithm-based fault tolerance for convolutional neural networks}.
\newblock \bibinfo{journal}{\emph{IEEE Transactions on Parallel and Distributed Systems}} \bibinfo{volume}{32}, \bibinfo{number}{7} (\bibinfo{year}{2020}), \bibinfo{pages}{1677--1689}.
\newblock


\end{thebibliography}

\clearpage
\appendix
\lstset{language=bash,
breaklines=true,
basicstyle=\ttfamily\footnotesize,
columns=fixed}
\section{Artifact Appendix}
This artifact contains instructions for how to replicate the experiments. It is available at:
\href{https://zenodo.org/records/14503617}{zenodo.org/records/14503617}.

\subsection{Evaluation environment}

\begin{itemize}
    \item Anaconda virtual environment with python 3.8.10
    \item CUDA Toolkit 12.6
    \item gcc 11.4.0
    \item Nividia A100 - 80GB
\end{itemize}

\subsection{Build and Config}
Choose one of the two build options
\subsubsection{Option 1: Docker Image}

Pull and run a pre-built docker image from Docker Hub.

\begin{lstlisting}[frame=single]
$ docker pull lyh911/attnchk-pytorch:2.0
$ docker run --ipc=host --shm-size=512m --gpus all -it --rm lyh911/attnchk-pytorch:2.0
\end{lstlisting}

\subsubsection{Option 2: Build from source}

1. Create an Anaconda environment.

\begin{lstlisting}[frame=single]
$ conda create --name attnchk python==3.8.10
$ conda activate attnchk
\end{lstlisting}


2. Download the source code.
\begin{lstlisting}[frame=single]
$ git clone https://github.com/liangyh911/ATTNChecker.git
$ cd ATTNChecker
\end{lstlisting}


3. Install the required Python packages for ATTNChecker.
\begin{lstlisting}[frame=single]
$ pip install -r requirements.txt
\end{lstlisting}

4. Move the modeling scripts to Huggingface Transformers package.

\begin{lstlisting}[frame=single]
$ cp ./HF-moding/modeling_bert.py <Path_To_Anaconda>/envs/attnchk/lib/python3.8/site-packages/transformers/models/bert/modeling_bert.py
$ cp ./HF-moding/modeling_gpt2.py      <Path_To_Anaconda>/envs/attnchk/lib/python3.8/site-packages/transformers/models/gpt2/modeling_gpt2.py
$ cp ./HF-moding/modeling_gpt_neo.py   <Path_To_Anaconda>/envs/attnchk/lib/python3.8/site-packages/transformers/models/gpt_neo/modeling_gpt_neo.py
$ cp ./HF-moding/modeling_roberta.py   <Path_To_Anaconda>/envs/attnchk/lib/python3.8/site-packages/transformers/models/roberta/modeling_roberta.py

\end{lstlisting}


5. Build PyTorch from the source.

\begin{lstlisting}[frame=single]
$ cd ./ATTNChecker/pytorch
$ conda install cmake ninja
$ pip install mkl-static mkl-include
$ pip install -r requirements.txt
$ git submodule sync
$ git submodule update --init --recursive
$ export CMAKE_PREFIX_PATH=${CONDA_PREFIX:-"$(dirname $(which conda))/../"}
$ python setup.py develop
\end{lstlisting}


If Python builds successfully, you will see in the last line of the output:
\begin{lstlisting}[frame=single]
Finished processing dependencies for torch==2.3.0a0+git83e4d29
\end{lstlisting}

Test if CUDA is available by using the following codes. $True$ will be returned if cuda is available.

\begin{lstlisting}[frame=single]
$ import torch
$ torch.cuda.is_available()
\end{lstlisting}

\subsection{Measuring ATTNChecker running overhead}

Use $./records/cleanRecords.py$ to clean the existing records before running new scripts.

\begin{lstlisting}[frame=single]
$ cd ./ATTNChecker
$ python ./records/cleanRecords.py
\end{lstlisting}

To measure the overhead of ATTNChecker, use the scripts in $./ABFT\_running\_time$ folder. The default settings are

\begin{itemize}
    \item batch size: 8
    \item number of training: 20
\end{itemize}

Before running the scripts, make sure to use one GPU if you have multiple GPUs on your device. Using export command.

\begin{lstlisting}[frame=single]
$ export CUDA_VISIBLE_DEVICES=0
\end{lstlisting}

Clean the previous records before running each test.

\begin{lstlisting}[frame=single]
# bert
$ python ./records/cleanRecords.py
$ python ./ABFT_running_time/bertTest.py
# gpt2
$ python ./records/cleanRecords.py
$ python ./ABFT_running_time/gpt2.py
# gpt neo
$ python ./records/cleanRecords.py
$ python ./ABFT_running_time/gpt-neo.py
# roberta
$ python ./records/cleanRecords.py
$ python ./ABFT_running_time/roberta.py
\end{lstlisting}

Here is an example output of the BERT model. For each test, the output results may vary. (The loss here is the 1-step training loss.)

\begin{lstlisting}[frame=single]
Attention Mechanism Overhead:  0.14775651309451615
Training Overhead:  0.0552227860653921
ATTNChecker Loss:  0.5106
no ATTNChecker Loss:  0.5106
\end{lstlisting}

\subsection{Measuring checkpoint save and load overhead}

Before measuring the save and load overhead of the checkpoint, please ensure you have tested ATTNChecker Running Overhead.

Use the scripts in $./Checkpoint\_time$ folder to test the save and load time of checkpointing a model. Here is an example.

\begin{lstlisting}[frame=single]
# bert
$ python ./records/cleanRecords.py
$ python ./Checkpoint_time/bert.py
# gpt2
$ python ./records/cleanRecords.py
$ python ./Checkpoint_time/gpt2.py
# gpt neo
$ python ./records/cleanRecords.py
$ python ./Checkpoint_time/gpt-neo.py
# roberta
$ python ./records/cleanRecords.py
$ python ./Checkpoint_time/roberta.py
\end{lstlisting}

Here is an example output of the BERT model. For each test, the output results may vary. The overhead is calculated in the same way as ATTNChecker running overhead.

\begin{lstlisting}[frame=single]
Overhead of Checkpointing:  8.522816600251735
\end{lstlisting}

\subsection{Measuring training loss of ATTNChecker and baseline during 3 epochs}
\begin{lstlisting}[frame=single]
# bert
$ python ./records/cleanRecords.py
$ python ./ABFT_epoch_loss/bert.py
# gpt2
$ python ./records/cleanRecords.py
$ python ./ABFT_epoch_loss/gpt2.py
# gpt neo
$ python ./records/cleanRecords.py
$ python ./ABFT_epoch_loss/gpt-neo.py
# roberta
$ python ./records/cleanRecords.py
$ python ./ABFT_epoch_loss/roberta.py
\end{lstlisting}

Here is an example output of the BERT model. For each test, the output results may vary. The baseline is training without ATTNChecker.

\begin{lstlisting}[frame=single]
Loss of ATTNChecker: 
1st epoch:  0.5349 , 2nd epoch:  0.3071 , 3rd epoch:  0.1285
Loss of Baseline: 
1st epoch:  0.5635 , 2nd epoch:  0.3362 , 3rd epoch:  0.1312    
\end{lstlisting}

\end{document}